\documentclass[final,english]{elsarticle}
\usepackage[T1]{fontenc}
\usepackage[latin9]{inputenc}
\usepackage{varioref}
\usepackage{units}
\usepackage{amsmath}
\usepackage{amssymb}
\usepackage{graphicx}
\usepackage{esint}
\usepackage{caption}
\usepackage{subcaption}
\usepackage{color}

\pdfpageheight\paperheight
\pdfpagewidth\paperwidth

\journal{Elsevier}
\usepackage{upgreek}
\usepackage[bookmarks,bookmarksopen,bookmarksdepth=6]{hyperref}
\usepackage{cases}
\usepackage{url}
\usepackage{lineno}

\usepackage{a4wide}

\usepackage{newtxtext,newtxmath,amsmath}
\makeatother

\usepackage{babel}
\begin{document}

\title{A computer program to simulate the response of SiPMs}

\author[]{E.~Garutti}
\author[]{R.~Klanner \corref{cor1}}
\author[]{J.~Rolph}
\author[]{J.~Schwandt}

\cortext[cor1]{Corresponding author. Email address: Robert.Klanner@desy.de,
 Tel.: +49 40 8998 2558}
\address{ Institute for Experimental Physics, University of Hamburg,
 \\Luruper Chaussee 147, D\,22761, Hamburg, Germany.}



 \begin{abstract}


 A Monte Carlo program which simulates the response of SiPMs is presented.
 Input to the program are the mean number and the time distribution of Geiger discharges from light, as well as the dark-count rate.
 For every primary Geiger discharge from light and dark counts in an event, correlated Geiger discharges due to prompt and delayed cross-talk and after-pulses are simulated, and a table of the amplitudes and times of all Geiger discharges in a specified time window generated.

 A number of different physics-based models and statistical treatments for the simulation of correlated Geiger discharges can be selected.
 These lists for many events together with different options for the pulse shapes of  single Geiger discharges are used to simulate charge spectra, as measured by a current-integrating charge-to-digital converter, or current transients convolved with an electronics response function, as recorded by a digital oscilloscope.
 The program can be used to compare simulations with different assumptions to experimental data, and thus find out which models are most appropriate for a given SiPM, optimise the operating conditions and readout for a given application or test programs which are used to extract SiPM parameters from experimental data.

 \end{abstract}


 \maketitle
 \tableofcontents
 \pagenumbering{arabic}

\section{Introduction}
 \label{sect:Introduction}

 SiPMs (Silicon Photo-MultiPliers), matrices of photo-diodes operated above breakdown voltage, are the photo-detectors of choice for many applications.
 They are robust, have single photon resolution, high photon-detection efficiency, operate at voltages below 100~V and are not affected by magnetic fields.
 However, their detection area is limited, their response is non-linear for high light intensities and their performance is affected by dark counts, prompt and delayed cross-talk and after-pulses.

 The simulation program which is described in this paper can be used to
 \begin{enumerate}
   \item test and verify analysis programs used to extract SiPM parameters from experimental data,
   \item determine SiPM parameters by comparing measurement and simulation results, and
   \item evaluate and understand the impact on the measurement of photons of different SiPM parameters, probability distributions for cross-talk and after-pulses and readout schemes.
 \end{enumerate}

 A number of simulation programs are documented in the literature.
 An incomplete list follows.
 A Monte Carlo program to simulates the multiplication process which is responsible for Geiger discharges is presented in Ref.~\cite{Spinelli:1997}.
 Programs which simulate the shape of the  transients for different options for the readout electronics are discussed in Refs.~\cite{Acerbi:2019, Calo:2019} and references therein.
 Monte Carlo programs addressing the readout of light from scintillators with SiPMs are documented in Refs.~\cite{Pulko:2012, Jha:2013, Niggemann:2015}, where the last one has been implemented in the GEANT4 framework~\cite{Dietz:2017}.
 The simulation discussed in Ref.~\cite{Gundacker:2013} puts the main emphasis on the optimisation of the time resolution using SiPMs for PET scanners.

 The Monte Carlo program described in this paper is less complete than some of the programs mentioned above, as neither the SiPM gain and photon-detection efficiency, nor the SiPM current response based on an electrical model are simulated.
 However, it is more flexible, as it allows to change
 \begin{itemize}
   \item the shape of the light pulses illuminating the SiPM,
   \item the pulse shape of single Geiger discharges,
   \item the probabilities and probability distributions for prompt and delayed cross-talk and after-pulses, and
   \item the readout mode and its electronic response function.
 \end{itemize}
 A number of different options are implemented in the program, and  extensions are straight-forward.

 The paper is structured in the following way.
 First, the parameters used to describe the SiPM performance and the different models implemented so far are discussed.
 Then, examples of the usage of the program are given: A comparison of simulation results with experimental data and studies of the impact of different assumption on measured charge spectra.
 Finally, the main results are summarized.

 \section{The SiPM response model}
  \label{sect:Model}

  First, an overview over the program flow is presented.
  The program simulates individual events.
  For every event primary Geiger discharges induced by light and by dark counts are generated, and their times and amplitudes are stored in an array, which is called \emph{Geiger Array}.
  Next, for every primary Geiger discharge time-correlated discharges due to prompt and delayed cross-talk and after-pulses are generated and their amplitudes and times are appended to the \emph{Geiger Array}.
  Electronics noise and random fluctuations for the individual Geiger discharges are taken into account.
  To every entry in the \emph{Geiger Array} a normalised current pulse with the shape of the current from a single Geiger discharge is assigned.

  The charge for an event is obtained by the sum of the integrals of the current pulses in the readout gate.
  For the current transients the time interval of the simulation is subdivided into time bins.
  The current transient of an event is obtained by the sum of the individual current transients in the time bins convolved with the response function of the readout.
  The 2-D distribution, pulse amplitude versus time difference for pulses exceeding a specified time interval, $\Delta t _{min}$, is obtained by time ordering the elements of the \emph{Geiger Array}, summing the pulses occurring in the time interval $\Delta t _{min}$ and calculating the time distance to the following pulse.
  As discussed in Ref.~\cite{Klanner:2019}, this 2-D distribution is a powerful tool to characterise SiPMs.

 In the following are presented the model parameters used, the assumptions for the different options implemented in the program and their physics motivation .
 More details on the functioning of SiPMs can be found in Refs.~\cite{Acerbi:2019, Calo:2019, Klanner:2019, Piemonte:2019}.
 Table~\ref{tab:Parameters} summarises the parameters and variables used.

 \emph{Time interval for the simulation and gate: $t _0$, $t _{start}$, $t _{gate}$, $n_t$.}
 The gate during which the current transient is integrated for the charge measurement, starts at $t_{start}$ and has the length $t_{gate}$.
 The simulation of  Geiger discharges is performed in the time interval  $-t_0 \leq t \leq t_{start} + t_{gate}$.
 The time interval $t_0 + t_{start}$, which precedes the start of the gate, has to be chosen sufficiently long so that the current integral of a Geiger discharge from a dark count at $t = -t_0$ is sufficiently small so that it can be ignored.
 The parameter $t _{start}$ is introduced so that delay curves can be simulated, which can be used to determine pulse shapes from charge measurements as discussed in Ref.~\cite{Chmill:2017}.
 For the simulation of the current transients the time interval between $t_0$ and $t_{start} + t_{gate}$ is divided in $n_t$ equal time bins.
 For $n_t$ a power of 2 is chosen so that the FFT (Fast Fourier Transform) algorithm with the number of operations increasing only $ \propto n_t \cdot \log (n_t )$ as opposed to $\propto n_t ^2$ can be used.

  \emph{Gain fluctuations: $\sigma_G$.}
  The gain of a SiPM is given by $ G = C \cdot (V_{bias} - V_{off} ) $, where
  $C$ is the sum of the capacitance of a single pixel and the capacitance parallel to the quenching resistor, and $V_{bias}$ the bias voltage and $V_{off}$ the voltage at which the Geiger discharge stops.
  It has been observed that differences in $V_{off}$ dominate the gain fluctuations~\cite{Chmill:2017}.
  The Geiger discharge stops when the voltage is too low to maintain the discharge.
  This is a statistical process which causes fluctuations of $V_{off}$.
  In addition, the average of $V_{off}$ can depend on the position of the Geiger discharge in a pixel because of variations of the electric field and may also be different from pixel to pixel. Last but not least also the capacitance $C$ may vary from pixel to pixel.
  In the simulation the mean amplitude of a primary Geiger discharge is set to one, and gain fluctuations are taken into account by multiplying the amplitudes of Geiger discharges by Gaussian-distributed random numbers with mean one and  $rms$ spread $\sigma_G$.

 \emph{Number, times and amplitudes of primary Geiger discharges from photons: $N_{\gamma G}$, $n_{\gamma G}$, $A _{i \gamma G} $, $t _{ i\gamma G}$.}
  $N_{\gamma G}$ is the mean number of primary Geiger discharges induced by photons, and $n_{\gamma G}$ the actual number for a given event.
  Implemented in the program are a fixed number, $n_ {\gamma G} = N_ {\gamma G}$, which assumes that $N_ {\gamma G}$ is an integer, or random $n_ {\gamma G}$ values generated according to a Poisson- or a Gauss-distribution.
  The Poisson distribution describes the statistics of a light pulser, like an LED or a laser, whereas the Gauss distribution is more appropriate for a calorimetric measurement.
  For the occurrence of the individual discharges, $i_{\gamma G}= 1 ~ ... ~ n_ {\gamma G} $, random times, $t _{i \gamma G}$, are generated.
  The choices offered by the program are
  a Gauss distribution, which approximates the light from an LED or a laser,
  the sum of two exponentials and
  the distribution $ (e^{-\lambda _1 \cdot t} - e^{-\lambda _2 \cdot t}) \cdot \lambda _1/(\lambda _2 - \lambda _1) $, which approximates the time distribution of photons from a scintillator, and from a scintillator read out via wave-length shifters.
  For the amplitudes, $A _{ i \gamma G} $, random numbers according to a Gauss distribution with mean one and $rms$~width $\sigma _G$ are generated.
  In this way fluctuations in the amplitudes of the Geiger discharges are taken into account.
  The values of $A _{ i \gamma G}$ and $t _{ i \gamma G} $ are stored in the \emph{Geiger array}, for the further analysis.

 \emph{Number, times and amplitudes of primary Geiger discharges from dark counts: $DCR$, $n_{DC}$, $A _{i DC }$, $t_{i DC}$.}
  Dark counts result from the thermal generation of electron-hole pairs which initiate Geiger discharges.
  They occur randomly in time, and their mean number in the time window of the simulation
  $N_{DC} = DCR \cdot (t_0 + t_{start} + t_{gate})$ for the dark-count rate $DCR$.
  The number of dark pulses in a given event, $n_{DC}$, is obtained as a Poisson-distributed random number with the mean $N_{DC}$.
  The time of an individual dark count, $t_{iDC}$, is obtained from a random number uniformly distributed between $-t_0$ and $t_{start} + t_{gate} $.
  As for the Geiger discharges from light, for the amplitudes, $A _{i DC}$, random numbers according to a Gauss distribution with mean one and $rms$~width $\sigma _G$ are generated.
  The list of $A _{i DC}$ and $t _{i DC}$ is appended to the \emph{Geiger Array}.

 \emph{Prompt cross-talk: $ p_{pXT}$, $A _{i pXT}$, $t _{i pXT}$.}
  Prompt cross-talk is caused by photons produced in the Geiger discharge which convert in the amplification region of a different pixel and cause a Geiger discharge there.
  Given the short light path, they can be considered  prompt.
  The light path can be directly through the silicon bulk, via reflection on the back or the front surface of the SiPM, or via a detector coupled to the SiPM.
  The introduction of trenches filled with light-absorbing material has significantly reduced the probability of prompt cross-talk~\cite{Acerbi:2019}.
  Three probability distributions for prompt cross-talk with the characteristic parameter $p_{pXT}$ are implemented: Binomial, Poisson and Borel\footnote{The Borel probability density function~\cite{Borel:1942} $e^{-p \cdot n} \cdot (p \cdot n)^{n-1}/n! $ describes the occurrence of $n$ events including the primary occurrence for the probability parameters $p$.
  Therefore the number of prompt cross-talk discharges using the Borel distribution is $n_{pXT} = n - 1$.
  In Ref.~\cite{Vinogradov:2012} it is shown that the Borel distribution can be used to describe the cross-talk distribution of SiPMs.}
  These distributions are used to generate for every primary Geiger discharge a random number, $n_{pXT}$, the number of prompt cross-talk Geiger discharges.
  If $n_{pXT} > 0$, the \emph{Geiger Array} is extended for every prompt Geiger discharge by $n_{pXT}$ entries with the times of the primary Geiger discharge and the amplitudes drawn from Gaussian-distributed random numbers with mean one and $rms$ width $\sigma _G $.

 \emph{Delayed cross-talk probability and time constant: $ p_{dXT},~\tau _{dXT}$, $A _{i dXT}$, $t _{i dXT}$.}
  Discharges which are delayed with respect to the primary Geiger discharge are called delayed cross-talk.
  The dominant source of delayed cross-talk are photons from the primary Geiger discharge which convert to an electron-hole pair in the non-depleted silicon.
  If the minority charge carrier diffuses into the amplification region, it can produce a delayed Geiger discharge there.
  Delayed cross-talk can be reduced by implementing special doping profiles between the amplification region and the non-depleted silicon.
  In the program only the delayed discharges in pixels other than the one with the primary Geiger discharge are included in the delayed cross-talk.
  They produce a current pulse with the same amplitude as the primary discharge.
  If the delayed discharge is in the same pixel as the primary Geiger discharge, the signal amplitude and the breakdown probability will be reduced because of recharging.
  As this situation cannot be distinguished from after-pulses it is included there.
  The simulation of delayed cross-talk is the same as for prompt cross-talk, with the exception that a delay time, $t_{dXT}$, is added to the time of the primary Geiger discharge.
  For $t_{dXT}$, a random number with the time distribution $e^{-t/\tau _{dXT}} \cdot \Theta(t) $ is generated, where $\Theta(t)$ is the Heavyside step function.
  As the origin of the delayed cross-talk are photons generating electron-hole pairs in the non-depleted bulk with  minority carriers diffusing into the multiplication region and producing Geiger discharges there, the exponential time dependence is certainly only a crude approximation.
  A more realistic model for the time dependence, if available, should be used.

 \emph{After-pulse probability, time constant and signal-recovery time: $ p_{AP}$, $\tau _{AP}$, $\tau _{rec}$, $A _{i AP}$, $t _{i AP}$. }
  After-pulses are delayed Geiger discharges in the same pixel in which the primary Geiger discharge has occurred.
  They can be caused by charges which are trapped by states in the silicon band gap and released after some time.
  Compared to delayed cross-talk, the simulation of after-pulses is more complicated as the probability as well as the amplitude are influenced by the preceding Geiger discharge.
  To simplify the simulation only a binomial distribution is used, which results in either no or one after-pulse.
  The time of a possible after-pulse, $t_{AP}$, is obtained from a random number with the time distribution $e^{-t/\tau _{AP}} \cdot \Theta(t) $, and the probability from $p_{AP} \cdot (1 - e^{-t_{AP}/\tau_{rec}} )$.
  The term $ (1 - e^{-t_{AP}/\tau_{rec}}) $ which takes into account that the Geiger-breakdown probability is reduced as long as the pixel is recharging, results in the reduced after-pulse probability $p _{AP} \cdot \tau _{rec} /(\tau _{rec} + \tau_{AP})$.
  For a similar reason, the average amplitude of the after-pulse is not one but $ (1 - e^{-t_{AP}/\tau_{s} } ) $, which is multiplied with a Gaussian-distributed random number with mean ona and $rms$ spread $\sigma _G$ to account for the gain fluctuations.

 \emph{Geiger Array: $n_G,~A_{iG},~t_{iG}$}.
  At this stage of the simulation the \emph{Geiger Array} contains for all $n_G$ Geiger discharges of an event the amplitudes, $A_{iG}$, and the time stamps, $t_{iG}$.
  This information is used in the following to calculate for a given pulse shape charge spectra, transients and time difference distributions.

 \emph{Pulse shape of a single Geiger discharge: $\tau _s ,~r_f,~\tau _f$. }
  Typical pulses from a SiPM have two components:
  A slow one from the recharging of the pixel, and a fast one if there is a capacitance in parallel with the quenching resistor~\cite{Klanner:2019}.
  For the pulse of a single Geiger discharge at $t=0$ the sum of two exponentials
  \begin{equation}\label{equ:I1}
    I_1 (t) = \Bigg(\frac{1-r_f}{\tau _s} \cdot e^{-t/\tau _s} + \frac{r_f}{\tau _f} \cdot e^{-t/\tau _f}\Bigg)\cdot \Theta(t),
  \end{equation}
  is implemented.
  The time constants of the fast and slow component are $\tau _f$ and $\tau _s$, respectively, and $r_f$ is the contribution of the fast component
  As the integral $\int _0 ^{\infty} I_1(t)~\mathrm{d}t = 1$, current amplitudes and charges use the same normalisation.

 \emph{Total charge and electronics noise: $Q,~\sigma_0$.}
  The  total charge of an event, $Q$, is obtained from the \emph{Geiger Array} by adding to
   \begin{equation}\label{equ:Qsum}
    Q' = \sum _{iG = 1} ^{n_G} A_{iG} \cdot \int _{t_{start}} ^{t_{start} + t_{gate}} I_1( t - t_{iG} )~ \mathrm{d}t
   \end{equation}
  a Gaussian-distributed random number with the mean 0 and the rms width $\sigma _0$, which takes into account the electronics noise.
  Looping over many events and entering the $Q$ values into a histogram, gives the final simulated charge spectrum.

 \emph{Current transient, electronics noise and response function: $n_t$, $I_{it}$, $t_{it}$, $\sigma _I$, $R_j$, $\sigma _R $.}
  Similar to Eq.~\ref{equ:Qsum}, the current transient for an event without electronics noise is
   \begin{equation}\label{equ:Isum}
    I'(t) = \sum _{iG = 1} ^{n_G} A_{iG} \cdot  I_1( t - t_{iG}),
   \end{equation}
  from which $I'_{it} = I'(t_{it})$, the current values at the times $t_{it}$, which have been introduced before, are obtained.
  Electronics noise is introduced by adding to every $I'_{it}$-value a Gaussian-distributed random number with mean 0 and $rms$ width $\sigma _I$.
  The result convolved with the electronics response function $R_{j}$ using the FFT (Fast Fourier Transform) gives $I_{it}$.
  Presently a Gauss function with an $rms$ width $\sigma _R$ is used for $R_j$.
  Other response functions including electronics filters can be easily implemented.

 \begin{table}[!ht]
  \centering
   \begin{tabular}{c||c|c}
     Group & Symbol & Description \\
    \hline \hline
                 & $\tau _s$ & time constant slow component \\ \cline{2-3}
     pulse shape & $\tau _f$ & time constant fast component \\ \cline{2-3}
                 & $r_f$     & fraction fast component \\
    \hline \hline
     gate        & $t_{start}$ & gate start \\ \cline{2-3}
     and         & $t_{gate}$ & gate length \\ \cline{2-3}
     time        & $-t_{0}$ & start time simulation \\ \cline{2-3}
     parameters  & $n_t$ & number time bins transient simulation \\
    \hline \hline
                      & $N_{\gamma G}$      & mean number discharges\\ \cline{2-3}
     primary Geiger   & $n_{ \gamma G}$     & actual number discharges\\ \cline{2-3}
     discharges light & $A_{i \gamma G}$    & amplitudes of discharges\\ \cline{2-3}
                      & $t_{i \gamma G}$    & times of discharges\\
    \hline \hline
           & $DCR$        & dark-count rate \\ \cline{2-3}
     dark  & $n_{DC}$     & actual number dark counts \\ \cline{2-3}
     counts& $A_{i DC}$ & amplitudes of discharges \\ \cline{2-3}
           & $t_{i DC}$ & times of discharges\\
    \hline \hline
                & $p_{pXT}$      & probability \\ \cline{2-3}
     prompt     & $n _{pXT}$     & actual number discharges \\ \cline{2-3}
     cross-talk & $A_{i pXT}$    & amplitudes of discharges \\ \cline{2-3}
                & $t_{i pXT}$    & times of discharges \\
    \hline \hline
                & $p_{dXT}$    & probability \\ \cline{2-3}
     delayed    & $\tau_{dXT}$ & time constant \\ \cline{2-3}
     cross-     & $n_{dXT}$    & actual number discharges \\ \cline{2-3}
     talk       & $A_{i dXT}$  & amplitudes of discharges \\ \cline{2-3}
                & $t_{i dXT}$  & times of discharges \\
    \hline \hline
                 & $p_{AP}$     & probability \\ \cline{2-3}
                 & $\tau_{AP}$  & time constant \\ \cline{2-3}
     after-      & $\tau_{rec}$ & Geiger probability recovery time \\ \cline{2-3}
     pulses      & $n_{AP}$     & actual number discharges \\ \cline{2-3}
                 & $A_{i AP}$   & amplitudes of discharges \\ \cline{2-3}
                 & $t_{i AP}$   & times of discharges \\ \cline{2-3}
    \hline \hline
     noise       & $\sigma_G$ & gain fluctuations \\ \cline{2-3}
     and         & $\sigma_0$ & electronics noise Q-measurement\\ \cline{2-3}
     electronic  & $\sigma_I$ & transient current noise \\  \cline{2-3}
     response    & $R(t)$     & electronic response function \\  \cline{2-3}
     function    & $\sigma_R$ & $rms$ width for Gaussian $R(t)$ \\
    \hline \hline
                  & $n_G$    & total number Geiger discharges \\ \cline{2-3}
     Geiger array & $A_{iG}$ & amplitudes of Geiger discharges \\ \cline{2-3}
                  & $t_{iG}$ & times of Geiger discharges \\
    \hline \hline
   \end{tabular}
    \caption{Parameters and variables used in the simulation program.
    \label{tab:Parameters} }
 \end{table}

 \section{Selected results}
  \label{sect:Results}

 \subsection{Time difference distribution for dark counts}
  \label{subsect:TimeDifference}

 The 2D distribution of the amplitude, $A$, versus the time difference, $\Delta t$, for dark counts is a powerful tool to characterise noise and correlated Geiger discharges of SiPMs.
 The time difference between consecutive Geiger discharges with a separation exceeding $\Delta t _{min}$ is denoted $\Delta t$.
 For $A$ the amplitudes of Geiger discharges with $\Delta t < \Delta t _{min}$ are added.
 The $A$~value of the later discharge or group of discharges is plotted.

  \begin{figure}[!ht]
   \centering
    \includegraphics[width=0.6\textwidth]{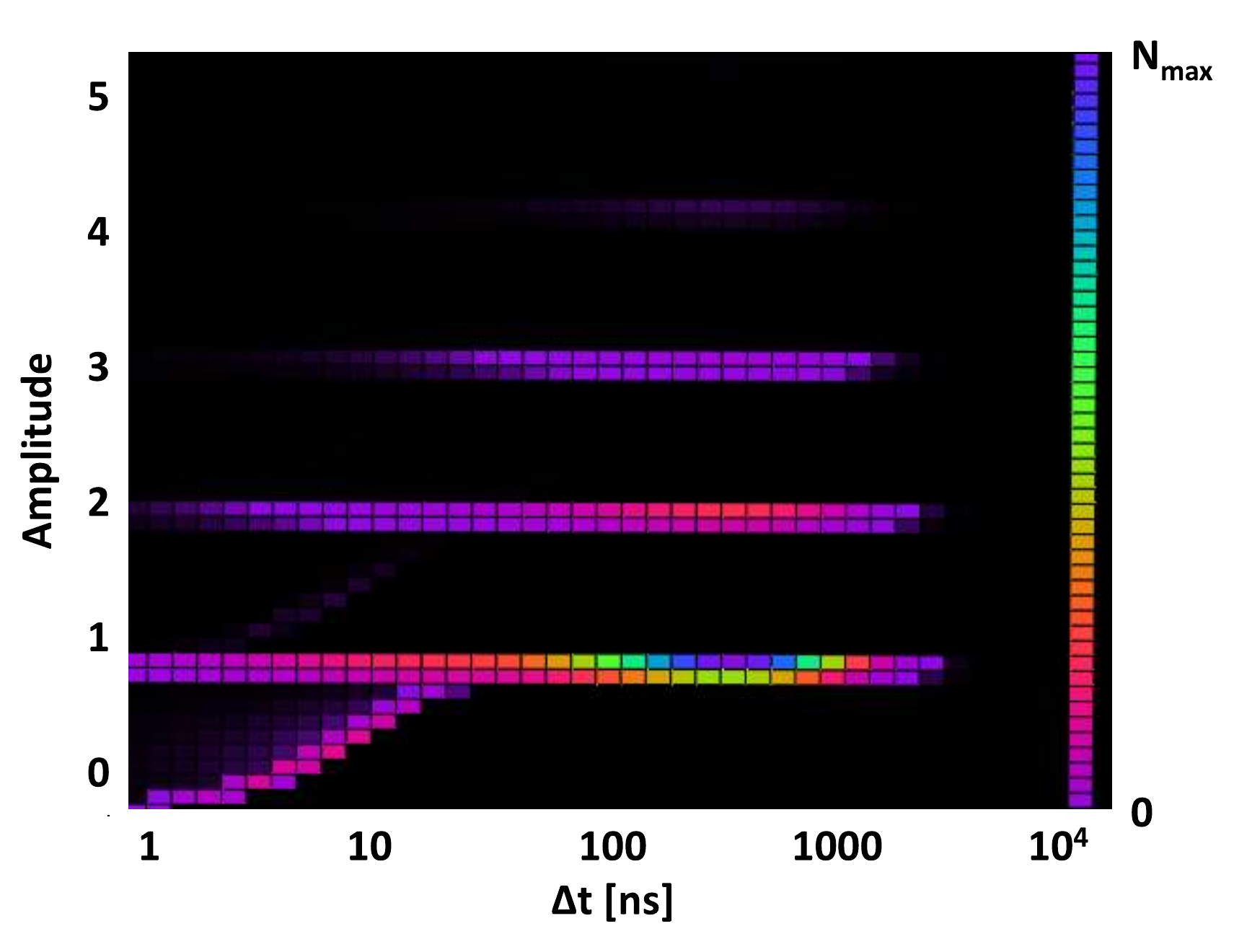}
   \caption{2D-distribution of simulated dark-count events in $\log(\Delta t) $~bins.
   $x$~axis:  time difference, $\Delta t$, between consecutive Geiger discharges;
   $y$~axis: current amplitude, $A$, of the later Geiger discharge.
   For more details see text (colour on-line).}
  \label{fig:2Ddt}
 \end{figure}

 Fig.~\ref{fig:2Ddt} shows for $ 5 \times 10^5$ events in a time interval of $50~\upmu$s for $DCR = 1$~MHz and $p_{pXT} = 0.2$, $p_{dXT} = 0.1$ and $p_{Ap} = 0.25$\footnote{These values of the parameters are significantly bigger than what is achieved by present day SiPMs. They have been chosen to better illustrate the different effects},
 the $A$ versus $\Delta t$ distribution in $\log ( \Delta t)$~bins.
 The horizontal bands for $A = 1$, 2, ... correspond to 1, 2, ... Geiger discharges within $\Delta t _{min}$.
 The rising band between $A = 0$ and 1 for $\Delta t \lesssim 50$~ns is due to after-pulses with reduced amplitudes because of the recharging of the pixel.

  \begin{figure}[!ht]
   \centering
   \begin{subfigure}[a]{0.5\textwidth}
    \includegraphics[width=\textwidth]{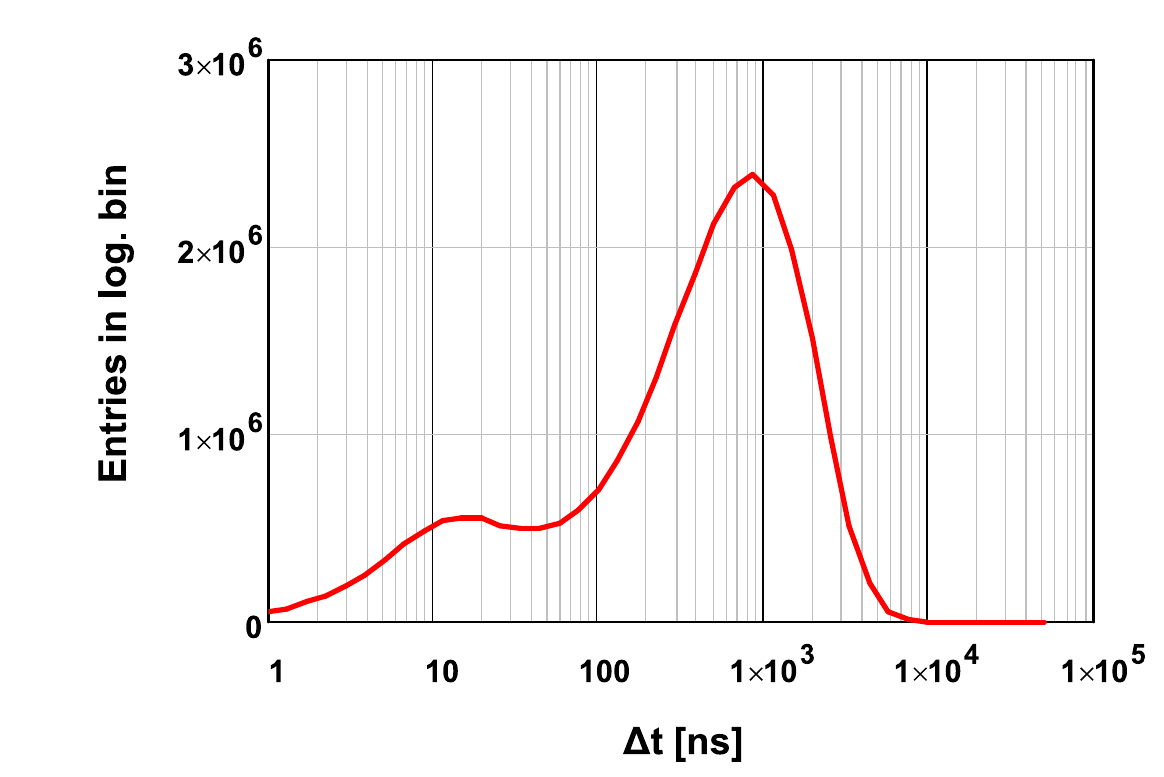}
    \caption{ }
    \label{fig:Logdt}
   \end{subfigure}%
   \begin{subfigure}[a]{0.5\textwidth}
    \includegraphics[width=\textwidth]{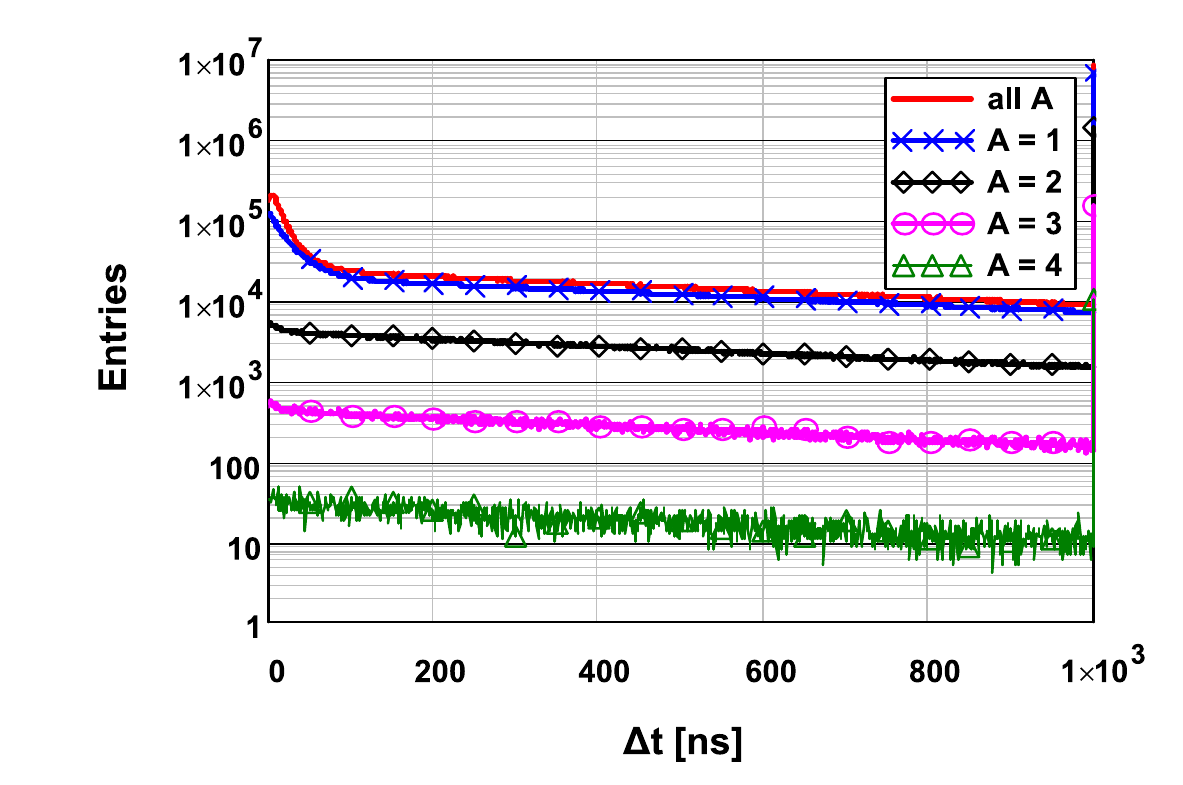}
    \caption{ }
    \label{fig:Lindt}
    \end{subfigure}%
   \caption{Distribution of the time difference, $\Delta t$,  between consecutive Geiger discharges separated by more than 1~ns for the $5 \times 10^5 $ simulated dark-count events presented in Fig.~\ref{fig:2Ddt}.
   (a) All Geiger discharges in $\log (\Delta t)$~bins.
   (b) Geiger discharges for  different selections of the amplitudes, $A$, in linear $\Delta t$~bins. }
   \label{fig:1Ddt}
  \end{figure}

 Fig.~\ref{fig:Logdt} shows the projection of the $A$ vs $\Delta t$ distribution on the $\Delta t$~axis in $\log ( \Delta t)$~bins.
 In the absence of correlated Geiger discharges the expected distribution for random dark counts is $\propto e^{- DCR \cdot \Delta t}$, which results for $\log ( \Delta t)$~bins in the dependence $ \Delta t \cdot e^{- DCR \cdot \Delta t}$.
 This explains the linear rise for $\Delta t \gtrsim 200$~ns and the peak close to $ 1 / DCR = 1 \upmu$s.
 The deviation from the linear dependence for $\Delta t \lesssim 200$~ns is due to correlated Geiger discharges from cross-talk and after-pulses.

 Fig.~\ref{fig:Lindt} shows the $\Delta t$~dependence in equal size bins for all $A$~values, and for the $A = 1$, 2,~... bands.
 For $\Delta t \gtrsim 200$~ns the expected dependence for dark counts $\propto e^{-DCR \cdot \Delta t}$ is observed for all $A$~values.
 For small $\Delta t$~values the curves deviate from the exponential dependence because of correlated Geiger discharges.
 The effect is strongest for \emph{all A}, where the increase towards $\Delta t _{min}$ is due to delayed cross-talk and after-pulses.
 The $A = 1$~increase is due to delayed cross-talk only.
 For $A \geq 2$ a delayed cross-talk causing an additional prompt cross-talk is required, which reduces the increase.

  \subsection{Comparison of charge spectra and transients with experimental data}
   \label{subsect:Comparison}

 In Ref.~\cite{Chmill:2017} detailed measurements for a KETEK SiPM with 4382 $15~\upmu \mathrm{m} \times 15~\upmu $m pixels are presented and a fit program is described and used to extract the SiPM parameters as a function of bias voltage.
 For this SiPM the turn-off voltage at room temperature is $V_{off} = 26.64~$V, and the pulse shape is a single exponential with the decay time $\tau \approx 20$~ns.
 For voltages between 29.5~V and 35.0~V in steps of 0.5~V, charge spectra with and without illumination by a pulsed LED and a gate of width  $t_{gate} = 100$~ns were recorded.
 For a fixed light intensity the average number of primary Geiger discharges, $N_{\gamma G}$, increases with voltage because of the increase of the Geiger probability.
 The values found from the fit program are $N_{\gamma G} = 0.79$ at 29.5~V increasing to 1.64 at 35~V.
 The dark-count rate, $DCR$, increases from 85~kHz to 210~kHz.
 For the voltage dependence of the other SiPM parameters we refer to Ref.~\cite{Chmill:2017}.

\begin{figure}[!ht]
   \centering
   \begin{subfigure}[a]{0.475\textwidth}
    \includegraphics[width=\textwidth]{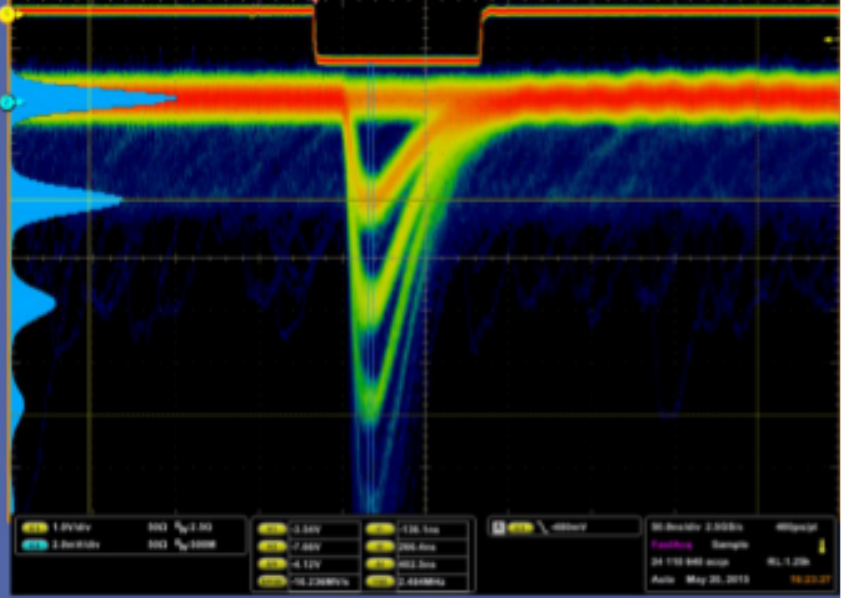}
    \caption{ }
    \label{fig:ScreenData}
   \end{subfigure}%
   \begin{subfigure}[a]{0.525\textwidth}
    \includegraphics[width=\textwidth]{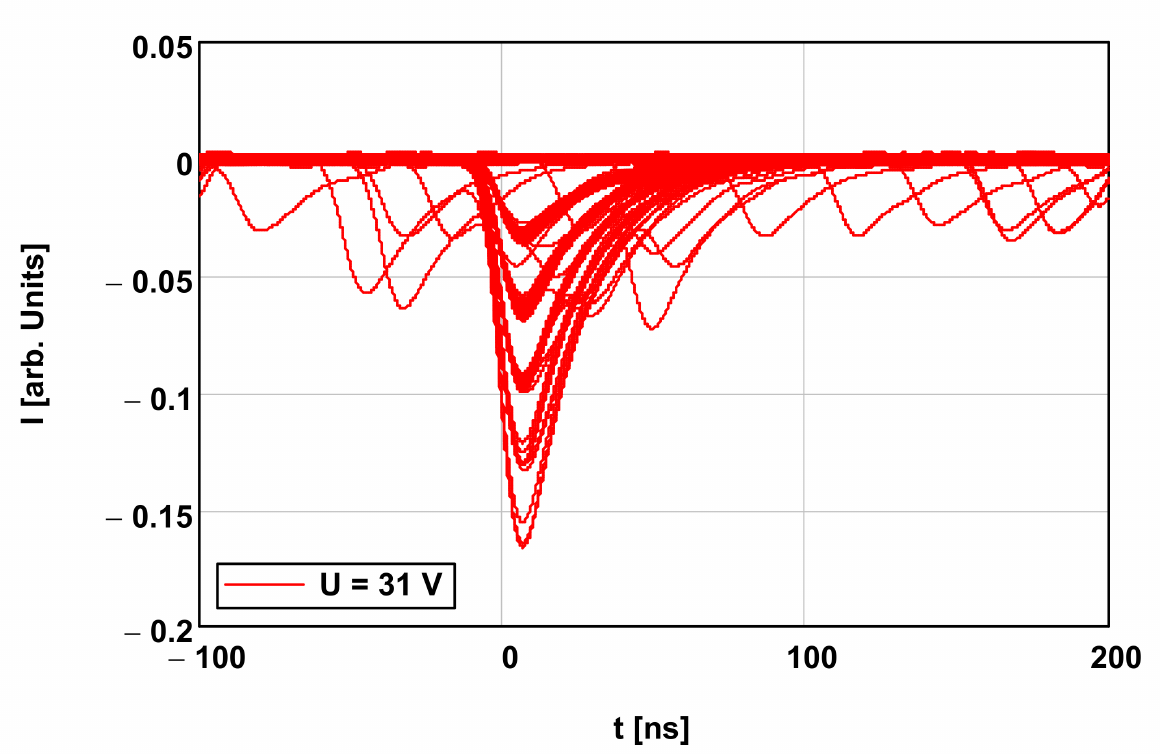}
    \caption{ }
    \label{fig:ScreenSim}
    \end{subfigure}%
   \caption{Comparison of (a) measured and (b) simulated transients at a 31.0~V bias voltage.
    The bands corresponding to zero, one and more Geiger discharges can be well distinguished.
    In addition, dark counts and delayed pulses are seen.
    The experimental data were recorded with the CAEN setup~\cite{Arosio:2014} which has a bandwidth of 125~MHz. }
   \label{fig:Screenshots}
  \end{figure}

 Fig.~\ref{fig:Screenshots} compares at 31~V the transient recorded with a CAEN setup (Ref.~\cite{Arosio:2014}) with 250 simulated transients.
 The comparison allows to estimate the pulse shape and the band width of the readout.

\begin{figure}[!ht]
   \centering
   \begin{subfigure}[a]{0.5\textwidth}
    \includegraphics[width=\textwidth]{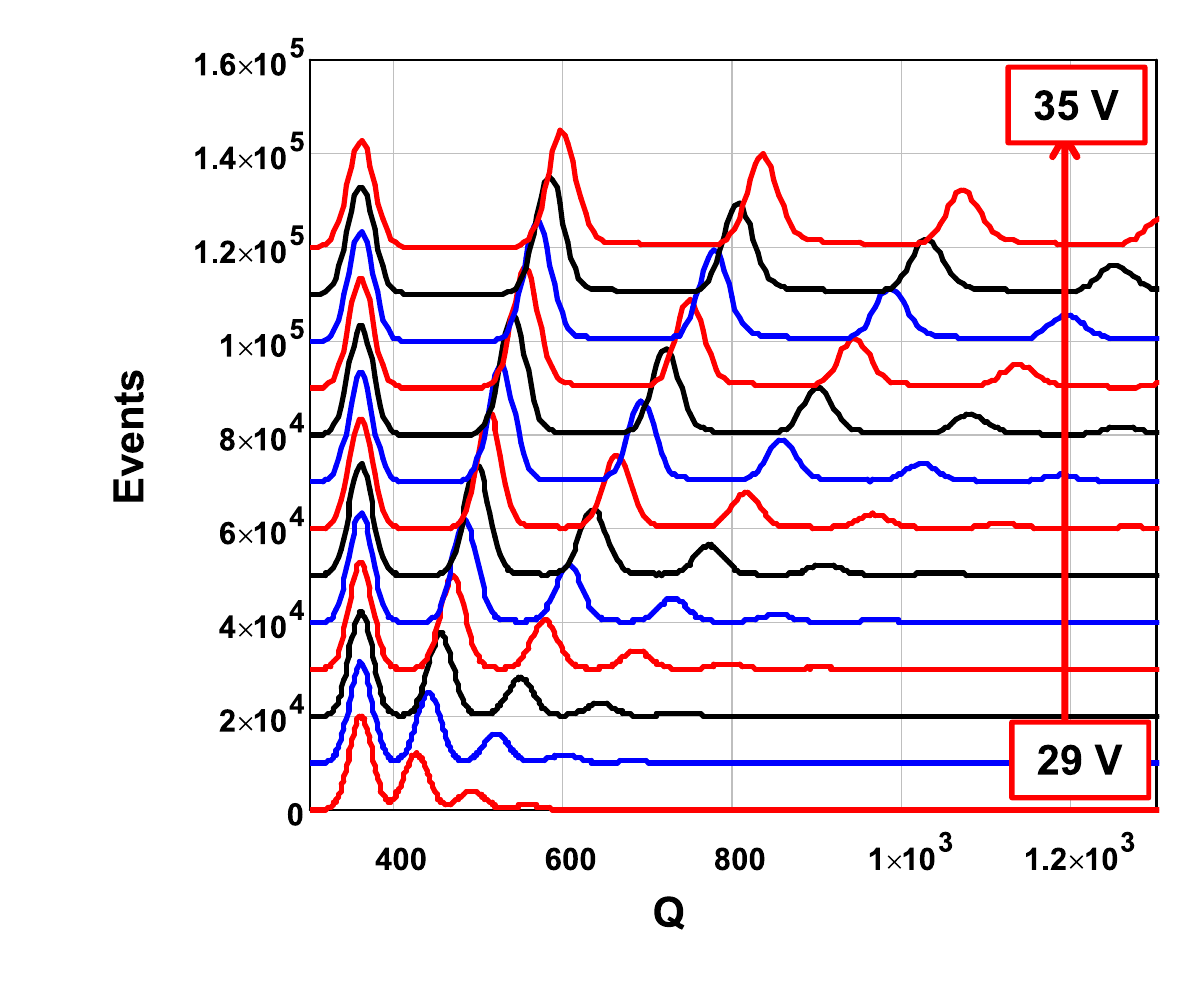}
    \caption{ }
    \label{fig:SimData_a}
   \end{subfigure}%
   \begin{subfigure}[a]{0.5\textwidth}
    \includegraphics[width=\textwidth]{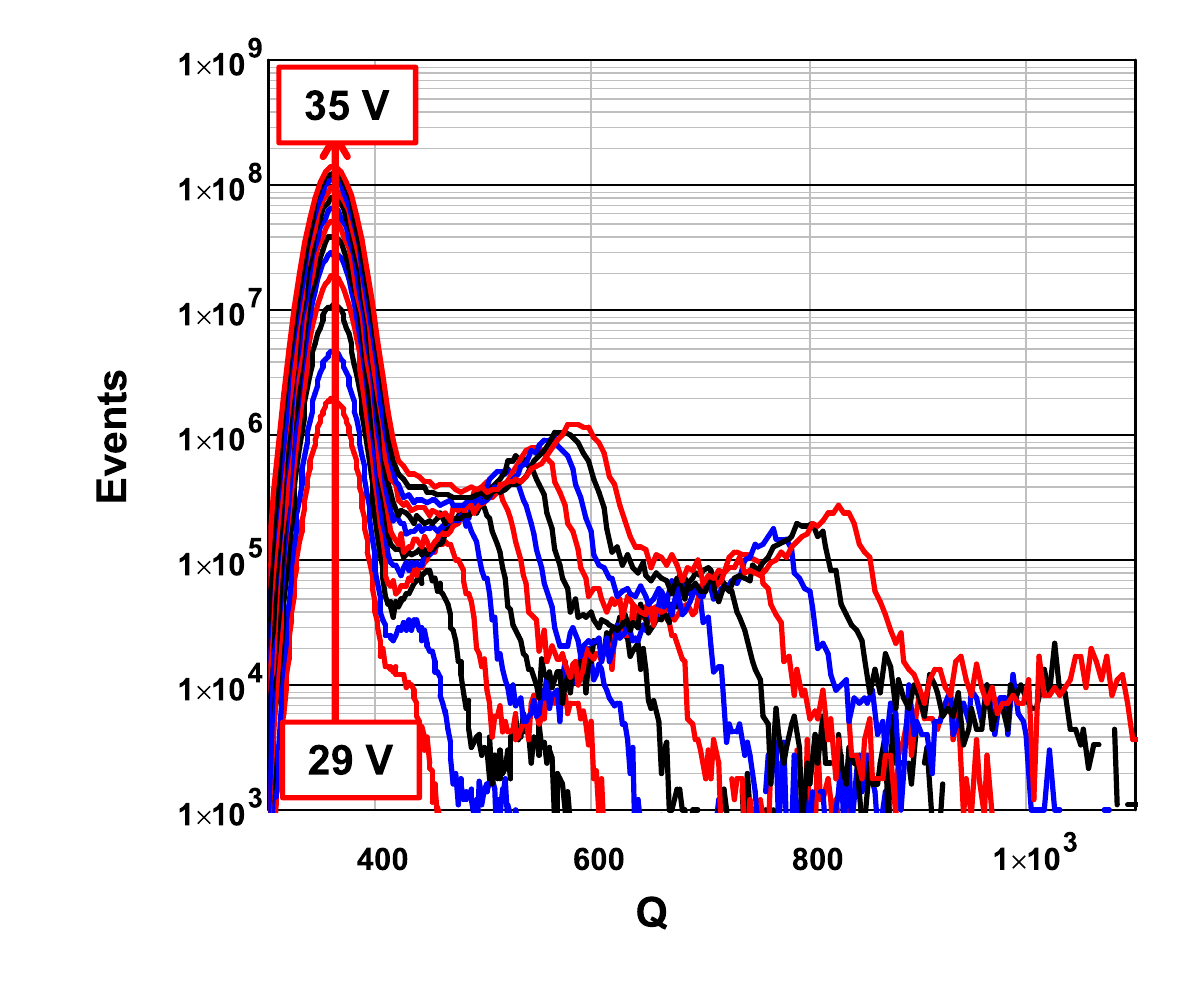}
    \caption{ }
    \label{fig:SimData_b}
   \end{subfigure}%

   \begin{subfigure}[a]{0.5\textwidth}
    \includegraphics[width=\textwidth]{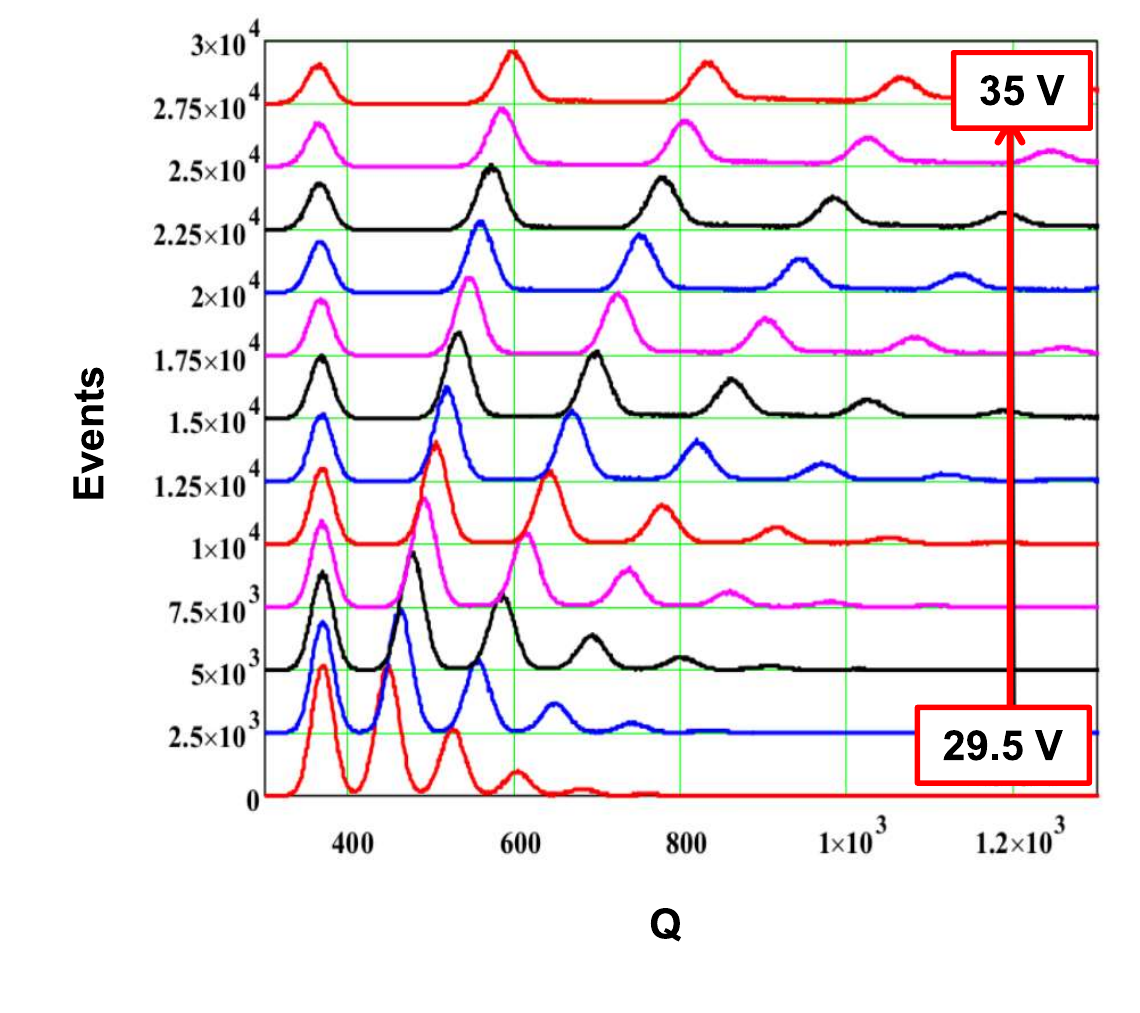}
    \caption{ }
    \label{fig:SimData_c}
   \end{subfigure}%
   \begin{subfigure}[a]{0.5\textwidth}
    \includegraphics[width=\textwidth]{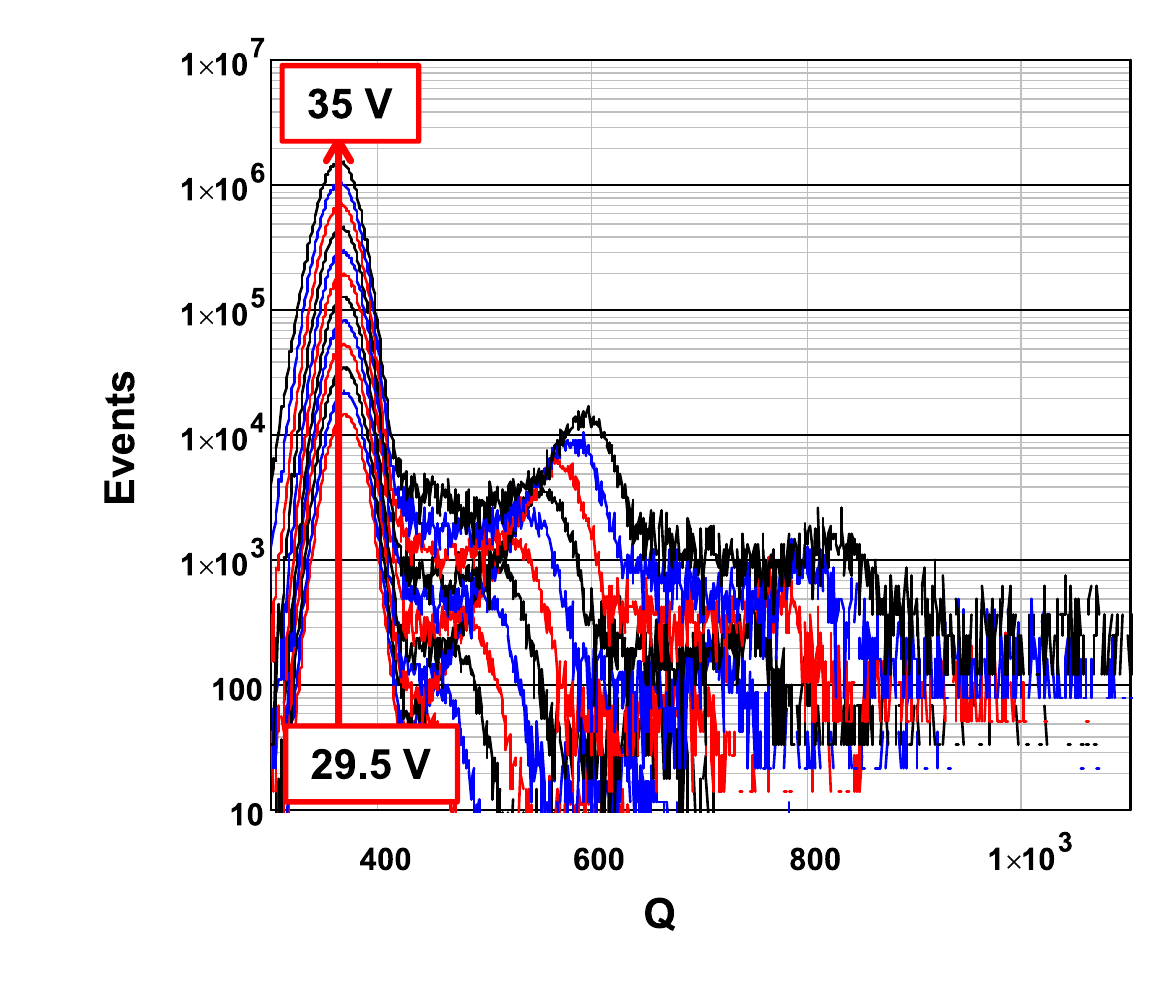}
    \caption{ }
    \label{fig:SimData_d}
   \end{subfigure}%

   \begin{subfigure}[a]{0.5\textwidth}
    \includegraphics[width=\textwidth]{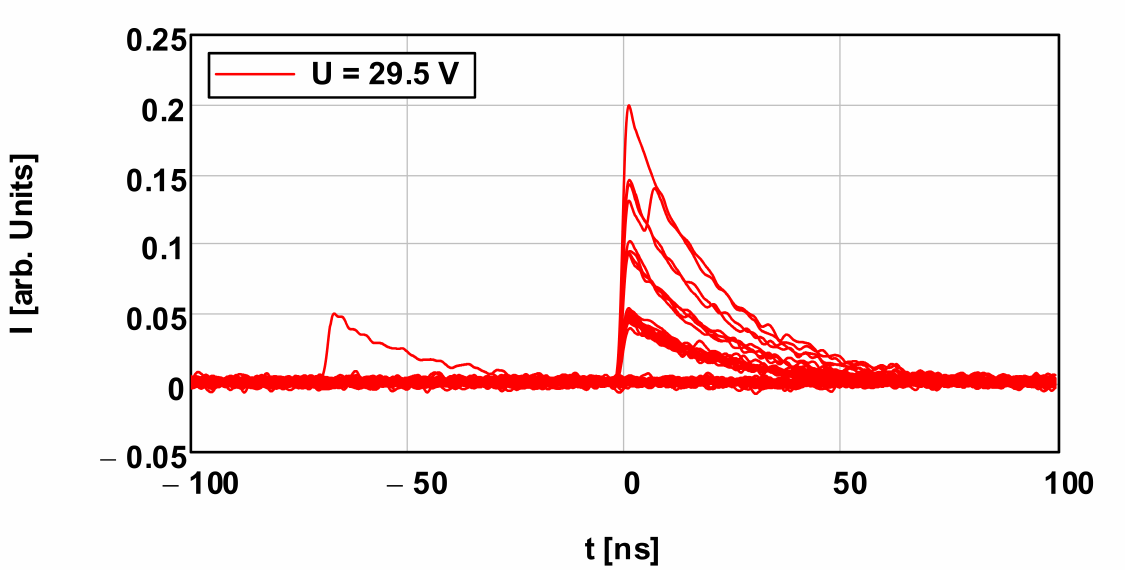}
    \caption{ }
    \label{fig:SimData_e}
   \end{subfigure}%
   \begin{subfigure}[a]{0.5\textwidth}
    \includegraphics[width=\textwidth]{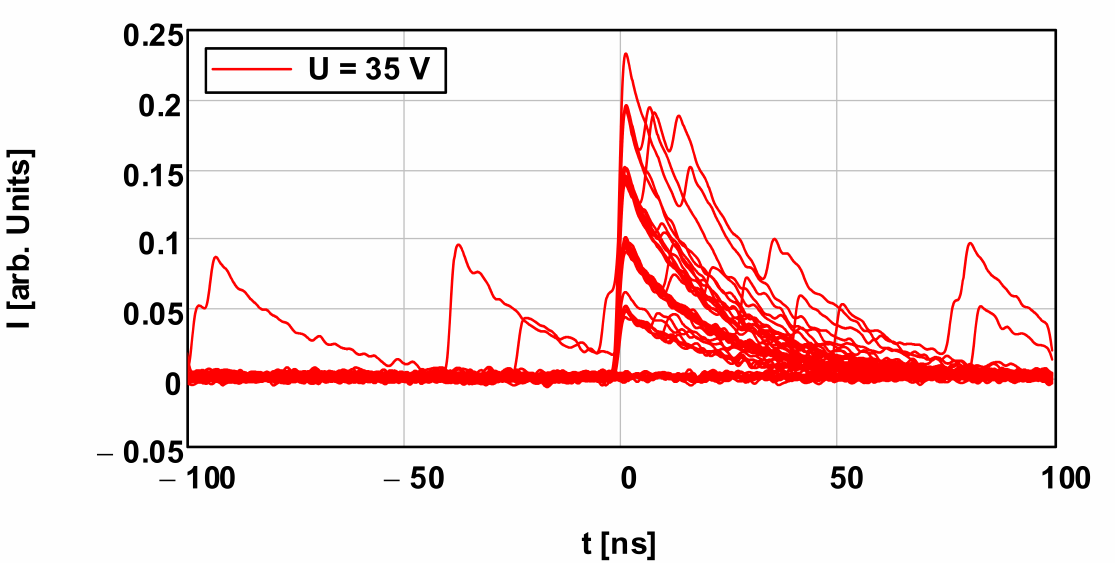}
    \caption{}
    \label{fig:SimData_f}
   \end{subfigure}%
   \caption{Comparison of simulated charge spectra with the experimental results of Ref.~\cite{Chmill:2017} for voltages between 29.5~V and 35.0~V:
    (a) Simulated charge spectra with illumination, shifted vertically by increasing multiples of $10^4$.
    (b) Simulated charge spectra without illumination, multiplied by an increasing power of 4.
    (c) Measured charge spectra with illumination, shifted by an increasing multiple of 2500.
    (c) Measured charge spectra without illumination, multiplied by an increasing power of 4.
    (d) 50 simulated transients at 29.5~V.
    (e) 50 simulated transients at 35.0~V.}
  \label{fig:SimData}
 \end{figure}

 Figs.~\ref{fig:SimData_a} and \ref{fig:SimData_c} show for the different voltages the simulated and  measured charge spectra with illumination, and Figs.~\ref{fig:SimData_b} and \ref{fig:SimData_d} without illumination.
 In order to allow a comparison between experimental data and simulations, the charge scale of the simulated data is scaled by the gain and shifted by the pedestal which were determined in Ref.~\cite{Chmill:2017}.
 A satisfactory qualitative agreement between measurements and simulations is found.
 Figs.~\ref{fig:SimData_e} and \ref{fig:SimData_f} show 50 simulated transients at 29.5~V and at 35~V, respectively.
 An increase in photon detection efficiency, in $DCR$, as seen from the pulses preceding the light pulse, and an increase in delayed cross-talk and after-pulses is clearly observed.

 For a more quantitative validation of the simulations, standard methods are used to extract SiPM parameters from the simulated charge spectra and compare them to the input values.

\begin{figure}[!ht]
   \centering
   \begin{subfigure}[a]{0.5\textwidth}
    \includegraphics[width=\textwidth]{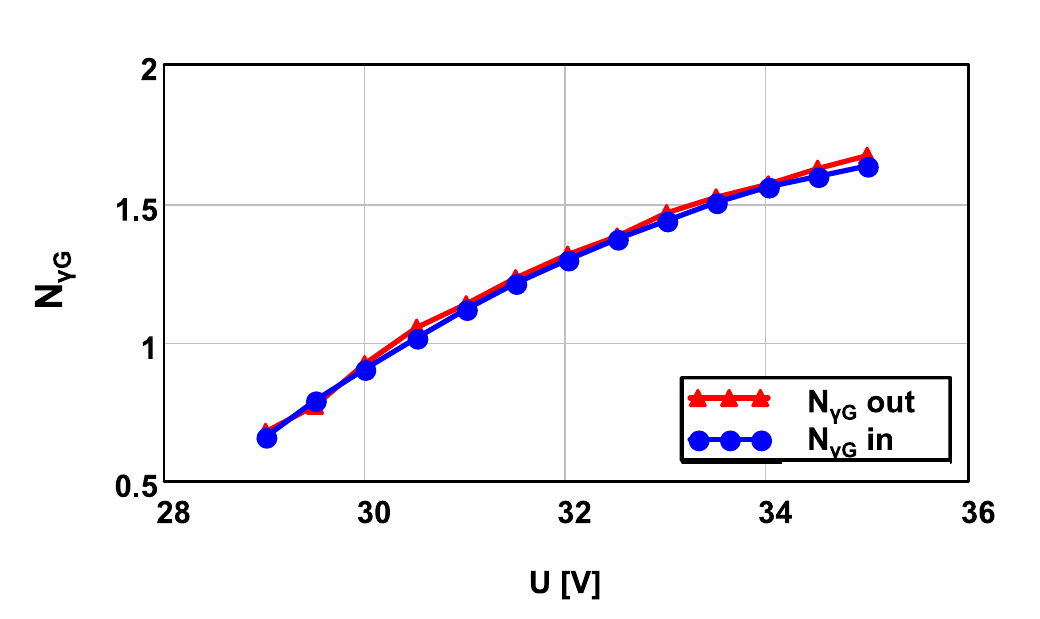}
    \caption{ }
    \label{fig:NpgV}
   \end{subfigure}%
   \begin{subfigure}[a]{0.5\textwidth}
    \includegraphics[width=\textwidth]{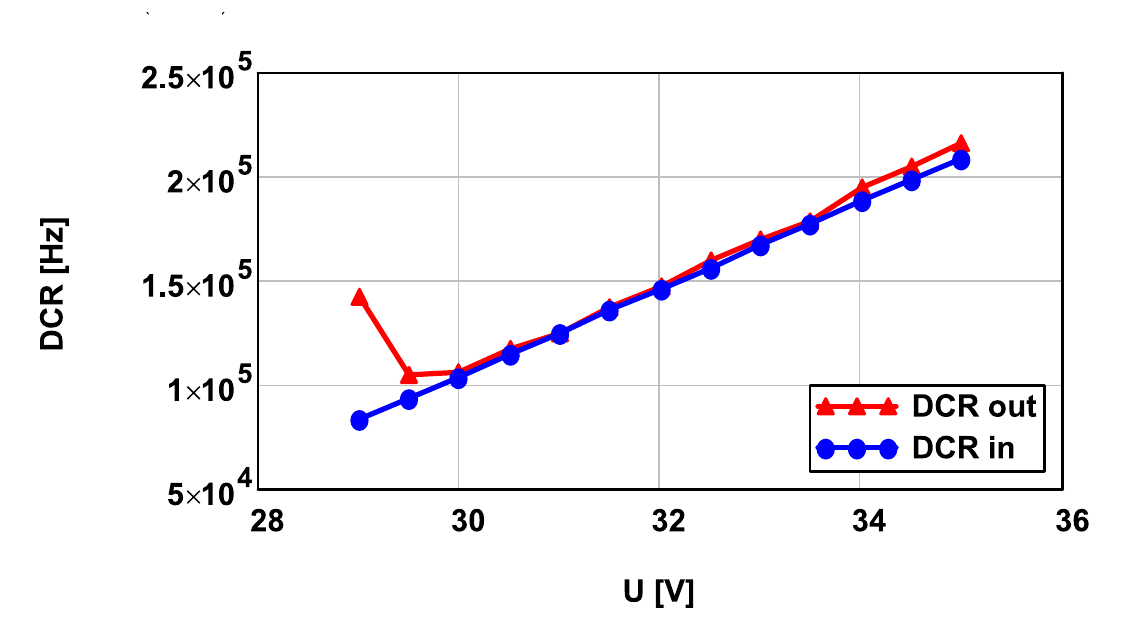}
    \caption{ }
    \label{fig:DCRV}
    \end{subfigure}%
   \caption{Comparison of the values of
    (a) the mean number of primary Geiger discharges from light, $N_{\gamma G}$, and
    (b) the dark-count rate, $DCR$,
   extracted from the simulated charge spectra (Figs.~\ref{fig:SimData_a}, \ref{fig:SimData_b}) with the input values obtained from the measured spectra (Figs.~\ref{fig:SimData_c}, \ref{fig:SimData_d}) in Ref.~\cite{Chmill:2017}.
   For more details see text. }
   \label{fig:NpgDCR}
  \end{figure}

 The mean number of primary Geiger discharges from the illumination, $N_{\gamma G}$, can be obtained from
 \begin{equation}\label{equ:NpG}
   N_{\gamma G} = -\ln\Big(N_{ped}^{\gamma }/N_{tot}^{\gamma }\Big) + \ln\Big(N_{ped}^{dark}/N_{tot}^{dark}\Big),
 \end{equation}
 with $N_{ped}^{\gamma }$ and $N_{ped}^{dark}$ the number of events in the pedestal with and without illumination, respectively, and $N_{tot}^{\gamma }$ and $N_{tot}^{dark}$ the corresponding total number of events.
 Fits of Gauss functions to the charge spectra in the pedestal regions are used to determine $N_{ped}$~values.
 Eq.~\ref{equ:NpG} uses the fact that in the absence of a primary Geiger discharge there is also no correlated noise, and that the probability of zero occurrences for a Poisson distribution with mean $N$ is $e^{- N}$.
 The second term in Eq.~\ref{equ:NpG} takes into account that dark counts reduce $N_{ped}^{\gamma }$.
 Fig.~\ref{fig:NpgV} compares as a function of voltage the $N_{\gamma G}$~values assumed in the simulation with the values obtained from the spectra of Fig.~\ref{fig:SimData} using Eq.~\ref{equ:NpG}.
 For all voltages the values agree within 1~\%.

 The dark-count rate, $DCR$, can be obtained from the spectra without illumination using
 \begin{equation}\label{eq:DCR}
   DCR = N_{> 0.5} ^{dark}/(N_{tot} ^{dark} \cdot t_{gate} )
 \end{equation}
 with $N_{> 0.5} ^{dark}$, the number of events in the dark spectrum above half the total charge of a single Geiger discharge.
 Fig.~\ref{fig:DCRV} compares as a function of voltage the $DCR$~values assumed in the simulation with the values obtained from the spectra of Fig.~\ref{fig:SimData_b}.
 For $U < 30$~V the peaks of 0 and 1 Geiger discharge are not well separated.
 As a result, the $DCR$~values determined are too high.
 In the region 30~V $\leq U \leq 32$~V the assumed and the extracted values agree within 1~\%.
 For higher voltages the extracted values are systematically higher than the input values.
 The deviation increases approximately linearly with voltage, and reaches a value of $\approx 3$~\% at 35~V.
 The reason is the increase of correlated Geiger discharges, which shift the charge to higher values compared to single Geiger discharges.
 As a result, pulses from dark counts with a partial overlap with the gate and charge $Q < 0.5$ move to $Q > 0.5$ in the presence of a correlated Geiger discharge.

 It is concluded that the simulation provides a realistic description of SiPM charge spectra and current transients, and thus can be used to study the influence of different parameters on the SiPM performance.

   \subsection{Influence of the fast component}
    \label{subsect:Fast}

\begin{figure}[!ht]
   \centering
   \begin{subfigure}[a]{0.5\textwidth}
    \includegraphics[width=\textwidth]{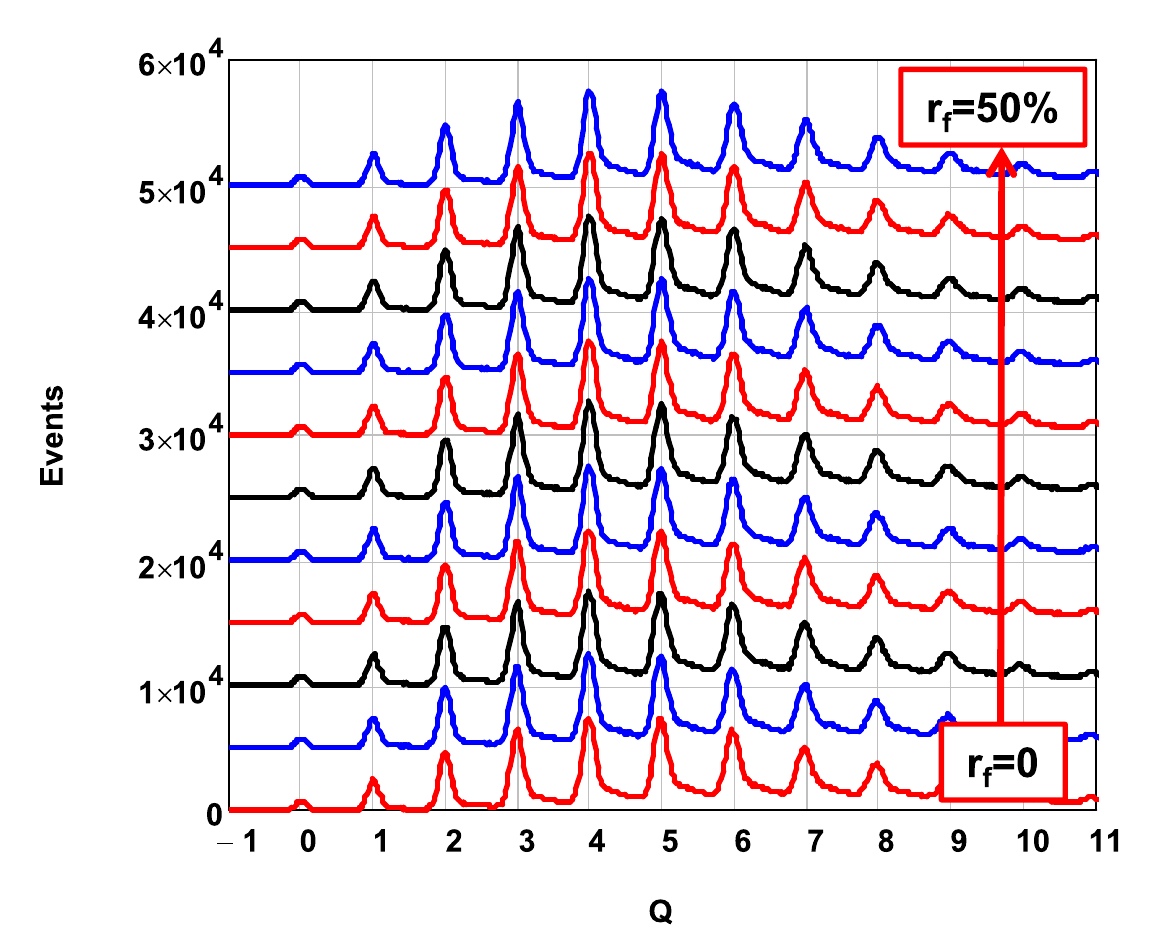}
    \caption{ }
    \label{fig:FastQ100ns}
   \end{subfigure}%
   \begin{subfigure}[a]{0.5\textwidth}
    \includegraphics[width=\textwidth]{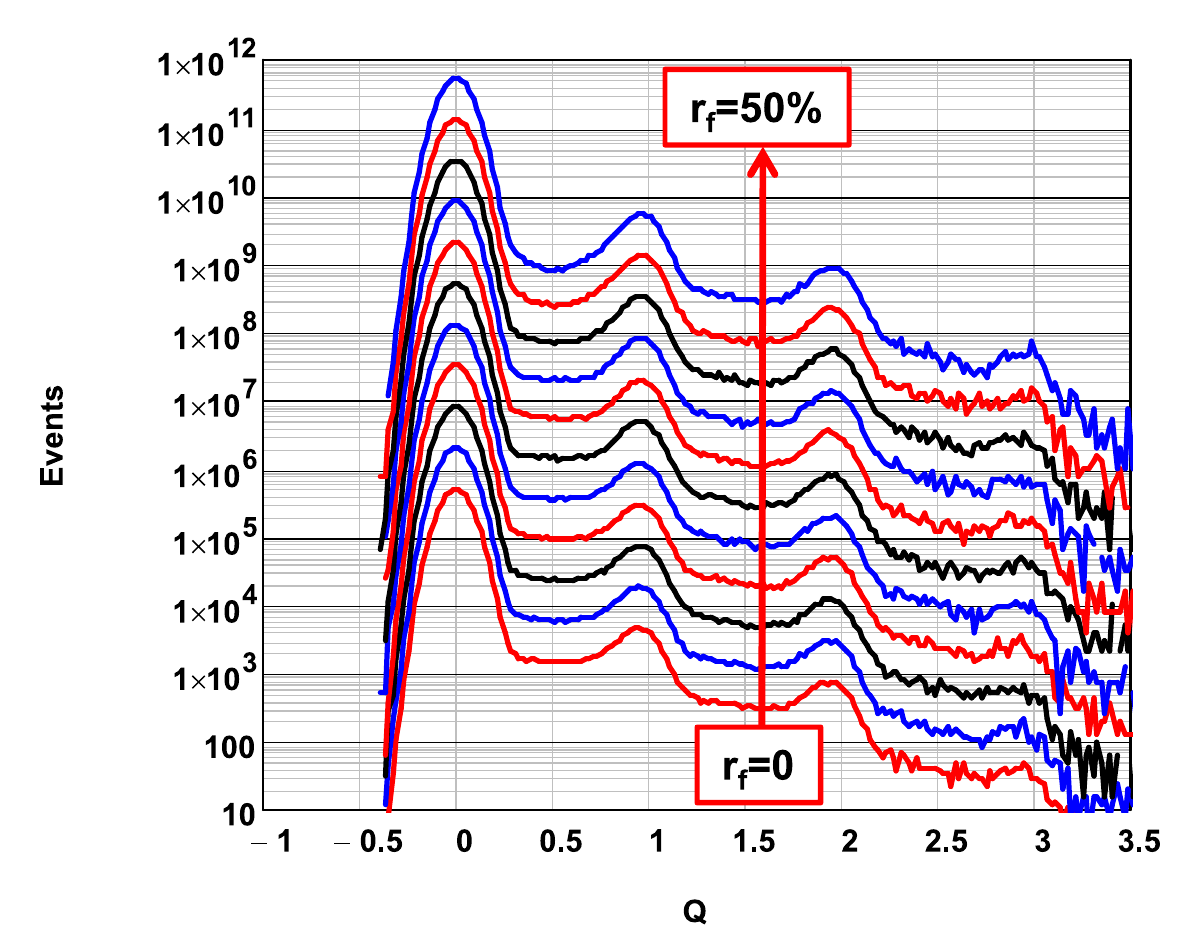}
    \caption{ }
    \label{fig:FastDC100ns}
   \end{subfigure}%
   \caption{Dependence of the simulated charge spectra on $r_f$, the contribution of the fast signal component, for $t_{gate} = 100$~ns and $DCR = 200$~kHz.
   (a) Illuminated SiPM with an average of 5 primary Geiger discharges.
   For clarity the spectra are shifted vertically by increasing multiples of 5000.
   (b) SiPM without illumination.
   For clarity the spectra are multiplied by increasing powers of 4.}
   \label{fig:Fast100ns}
  \end{figure}

\begin{figure}[!ht]
   \centering
   \begin{subfigure}[a]{0.5\textwidth}
    \includegraphics[width=\textwidth]{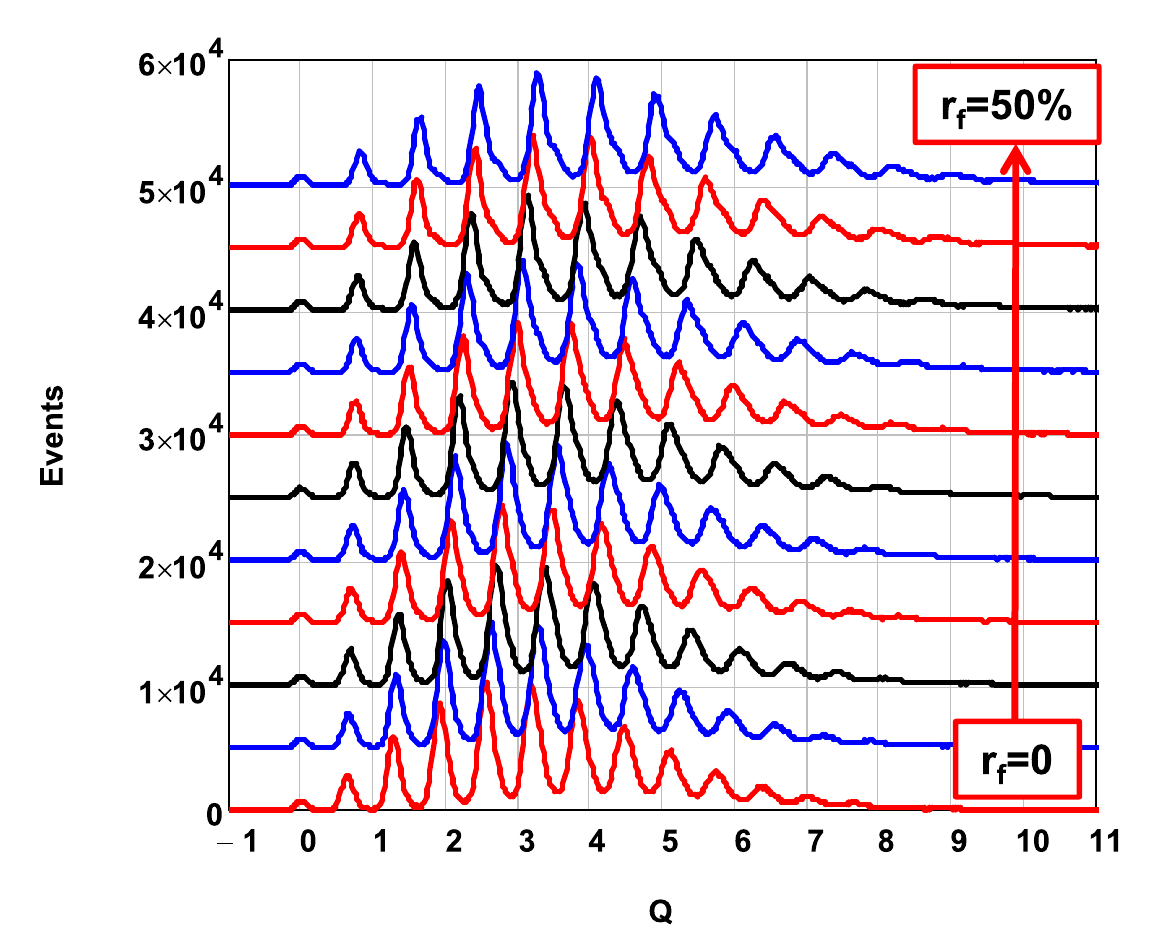}
    \caption{ }
    \label{fig:FastQ20ns}
   \end{subfigure}%
   \begin{subfigure}[a]{0.5\textwidth}
    \includegraphics[width=\textwidth]{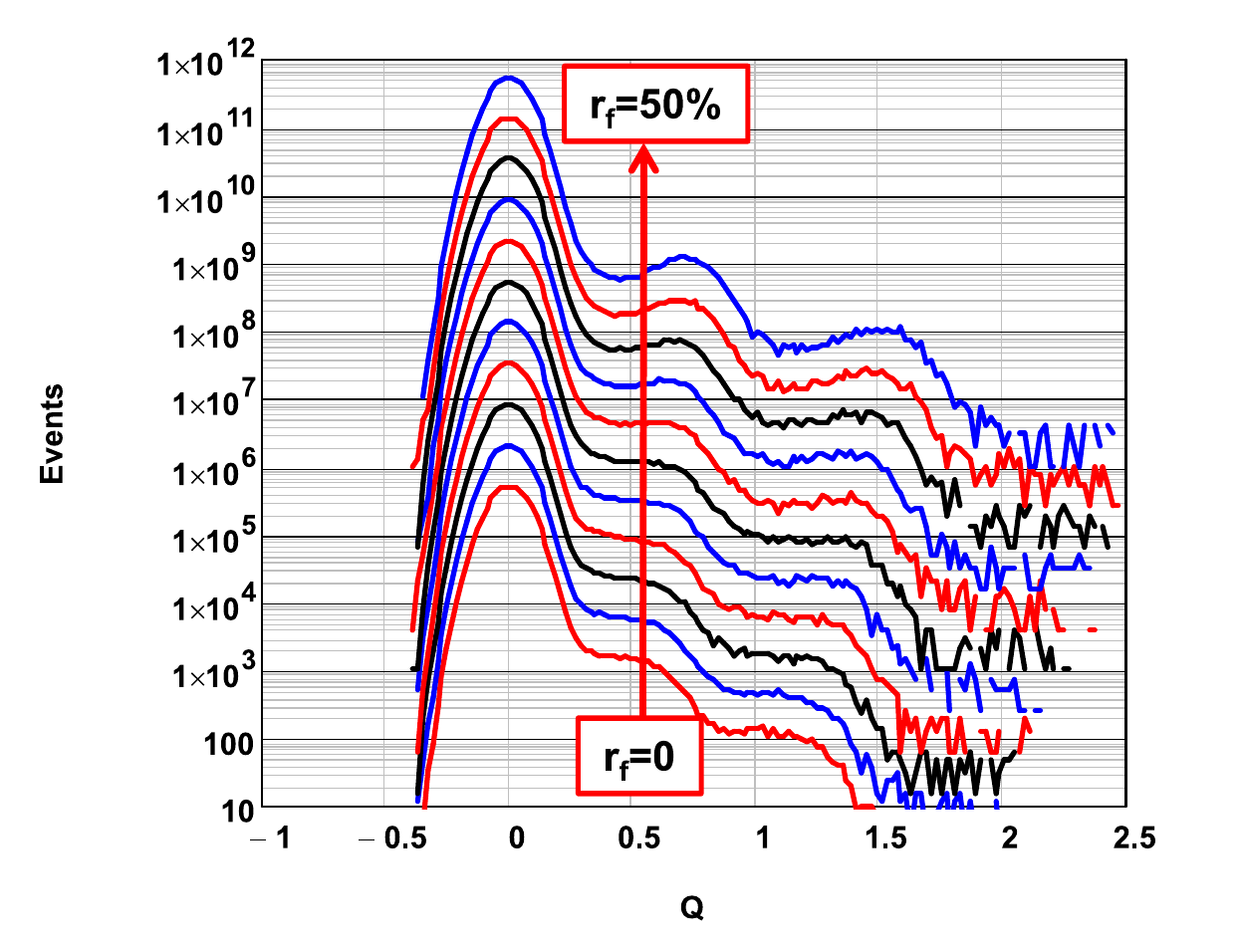}
    \caption{ }
    \label{fig:FastDC20ns}
    \end{subfigure}%
   \caption{Same as Fig.~\ref{fig:Fast100ns}, however for $t_{gate} = 20$~ns.}
   \label{fig:Fast20ns}
  \end{figure}

 Figs.~\ref{fig:Fast100ns} and \ref{fig:Fast20ns} show the simulated charge spectra for the SiPM with and without illumination for $r_f$, the fraction of the fast component, between 0 and 50~\% in 5~\% steps and $t_ {gate}$ values of 100~ns and 20~ns.
 The time constants of the fast and slow component of the signal are 1.5~ns and 20~ns, respectively.
 The charge $Q$ is normalised so that the integral of the current of a single Geiger discharge (Eq.~\ref{equ:I1}), $\int _0 ^\infty I_1(t)~\mathrm{d}t = 1$.
 The parameters chosen for the simulation are $DCR = 200$~kHz for the dark-count rate, and $N_{\gamma G} =5$ for the mean number of primary Geiger discharges from the illumination.

 Fig.~\ref{fig:FastQ100ns} shows that the peak positions of the charge spectra for $t_ {gate} = 100$~ns with illumination are not affected by a change of $r_f$.
 The reason is that both the fast and the slow component of the prompt Geiger discharges are fully integrated.
 However, the region in between the peaks is affected by $r_f$.
 The number of entries at $Q = 0.5$ decreases by a factor 2 between $r_f = 0$ and $r_f = 50$~\%.
 This region is populated by dark counts occurring approximately one time constants before the start and before the end of the gate, resulting in $Q \approx 0.5$.
 For $\tau _f \ll \tau _s$, the fast component hardly contributes and the fraction of events at $Q \approx 0.5$ decreases approximately $\propto (1-r_f)$.

 As expected the results for  $N_{\gamma G}$, as obtained from the fraction of events in the pedestal peak, are not affected by the change of $r_f$.
 The same holds for the extraction of the dark-count rate from the spectra of Fig.~\ref{fig:FastDC100ns}.
 Similar to Sect.~\ref{subsect:Comparison} $DCR = 205$~kHz is obtained, which is again a few percent higher than the input value of 200~kHz.

 Fig.~\ref{fig:FastQ20ns} shows that the peak positions of the charge spectra for $t_ {gate} = 20$~ns with illumination increase as a function of $r_f$.
 The peak from a single Geiger discharge moves from $\approx 0.5$ for $r_f = 0$ to 0.8 for $r_f = 50$~\%.
 The reason is that the entire fast component but only a fraction of the slow component are integrated by the short gate.
 The determination of $N_{\gamma G}$ is not affected by $r_f$.
 The dark spectra of Fig.~\ref{fig:FastDC20ns} for $t_ {gate} = 20$~ns show that for $r_f = 0$ only a shoulder appears at the position of one prompt Geiger discharge, which develops into a peak with increasing $r_f$.
 As a result the determination of $DCR$ using the method described in  Sect.~\ref{subsect:Comparison} becomes unreliable for shorter gates.

   \subsection{Influence of dark counts}
    \label{subsect:DC}

 In this section the influence of the dark-count rate, $DCR$, on the charge spectra for two different gate lengths, $t_{gate}$, is investigated.
 This is relevant for the use of SiPMs in high radiation environments, which causes an increase in $DCR$ by radiation damage~(Ref.~\cite{Garutti:2019}, \cite{Cerioli:2020}), and for applications in the presence of background light.

\begin{figure}[!ht]
   \centering
   \begin{subfigure}[a]{0.5\textwidth}
    \includegraphics[width=\textwidth]{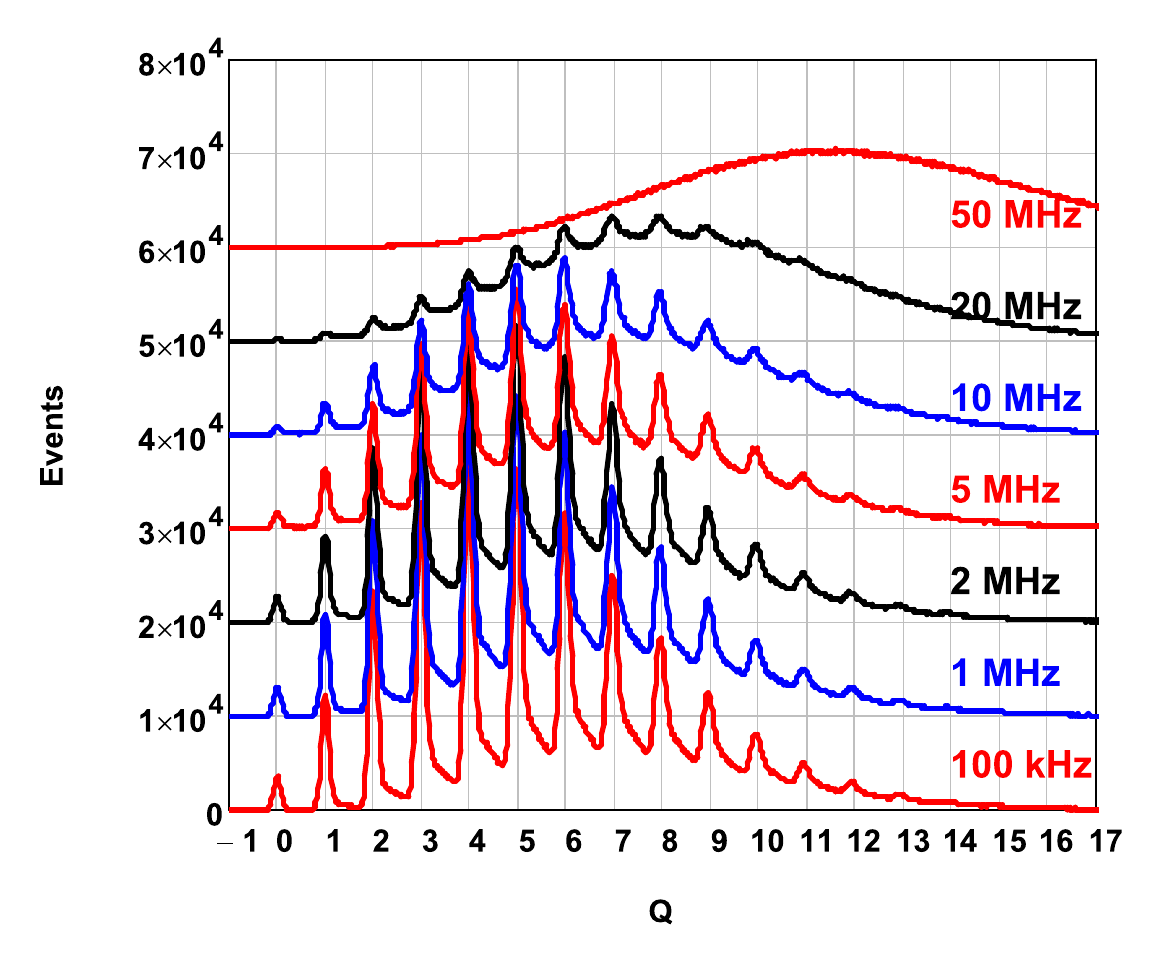}
    \caption{ }
    \label{fig:QDCR100ns}
   \end{subfigure}%
   \begin{subfigure}[a]{0.5\textwidth}
    \includegraphics[width=\textwidth]{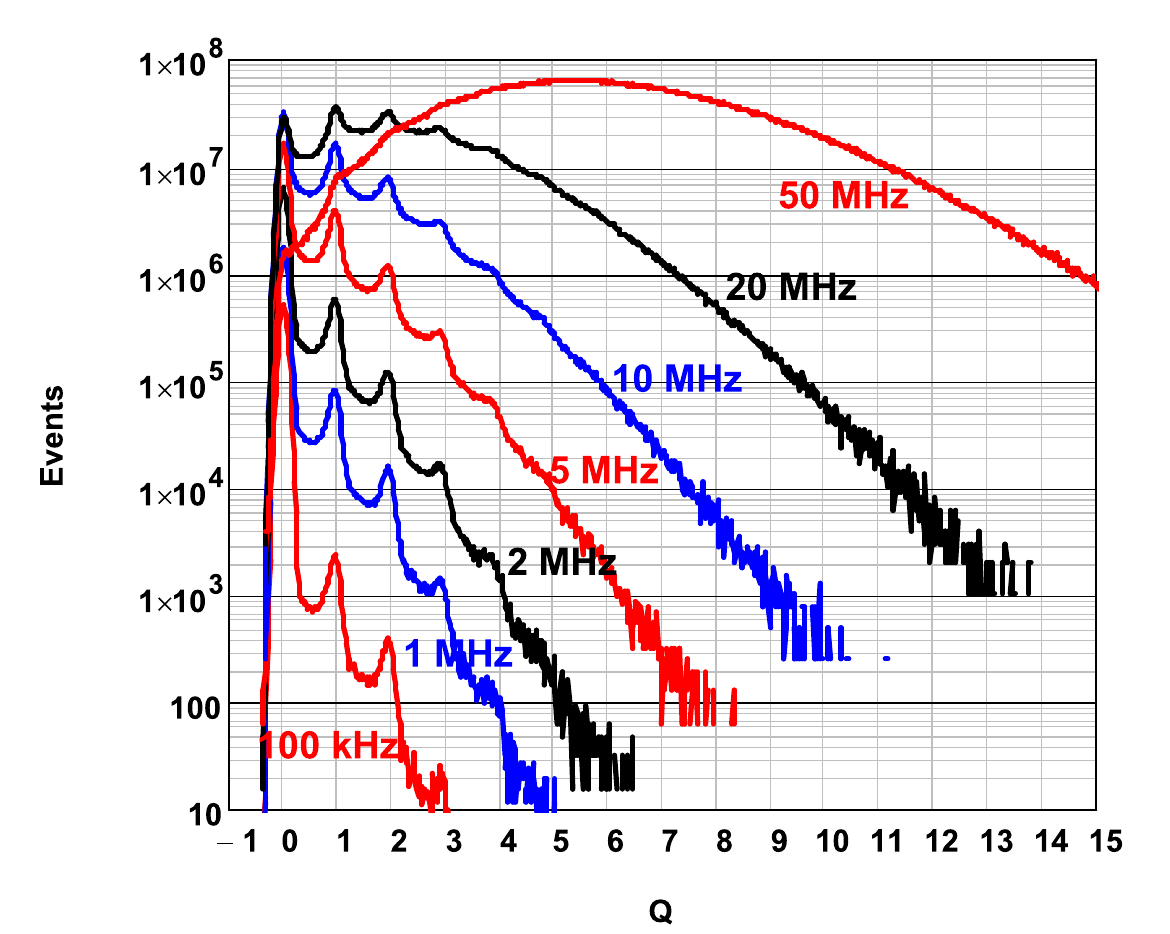}
    \caption{ }
    \label{fig:DCDCR100ns}
    \end{subfigure}%
   \caption{Simulated charge spectra as a function of the dark-count rate, $DCR$, between 100~kHz and 50~MHz for pulses with the time constant of $\tau _s = 20$~ns and a gate length $t_{gate} = 100$~ns.
   (a) Illuminated SiPM with a mean number of primary Geiger discharge $N_{\gamma G} = 5$.
       For clarity the spectra are shifted vertically by an increasing multiple of $10^{4}$.
   (b) SiPM without illumination.
       For clarity the spectra are multiplied by an increasing power of 4.}
   \label{fig:QvsDCR100ns}
  \end{figure}

\begin{figure}[!ht]
   \centering
   \begin{subfigure}[a]{0.5\textwidth}
    \includegraphics[width=\textwidth]{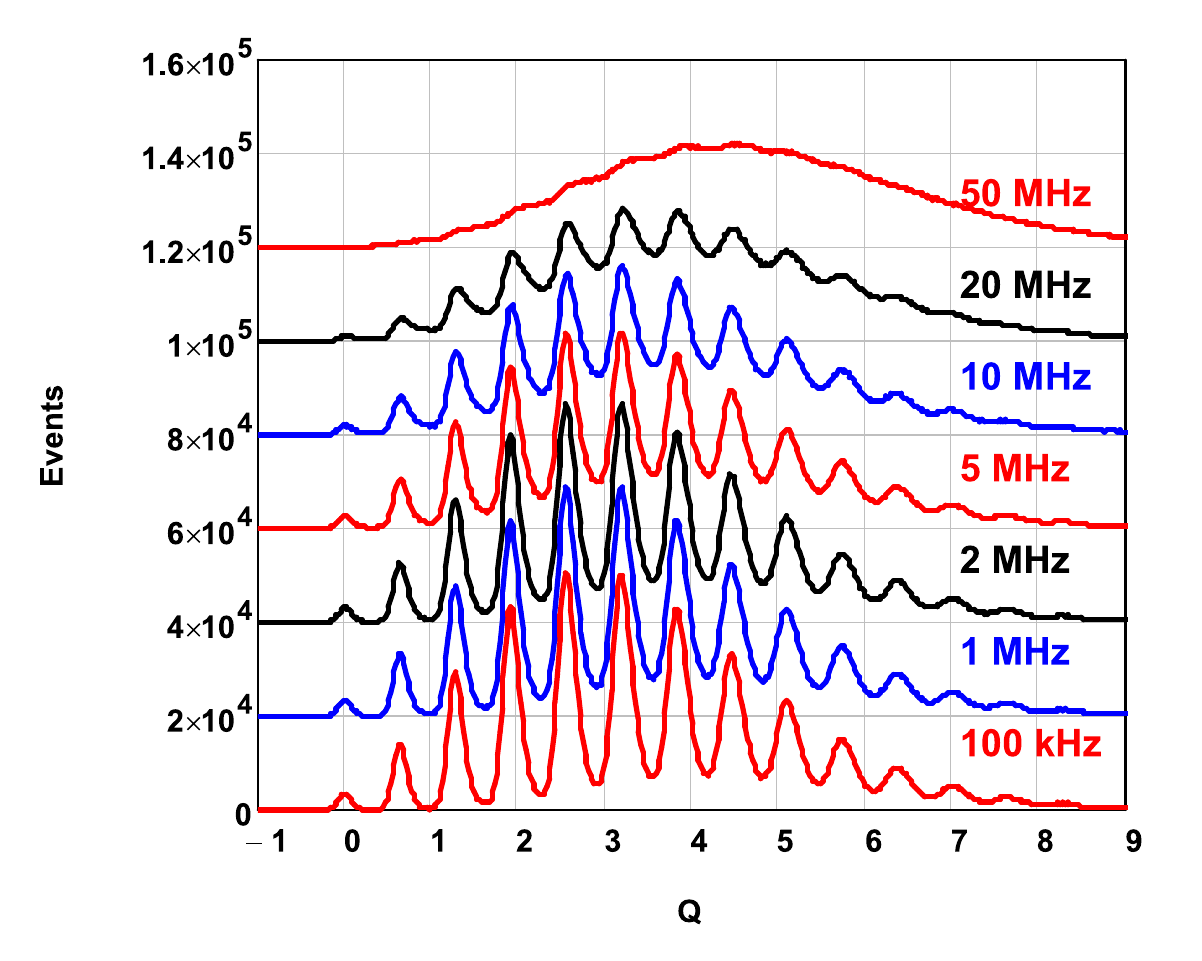}
    \caption{ }
    \label{fig:QDCR20ns}
   \end{subfigure}%
   \begin{subfigure}[a]{0.5\textwidth}
    \includegraphics[width=\textwidth]{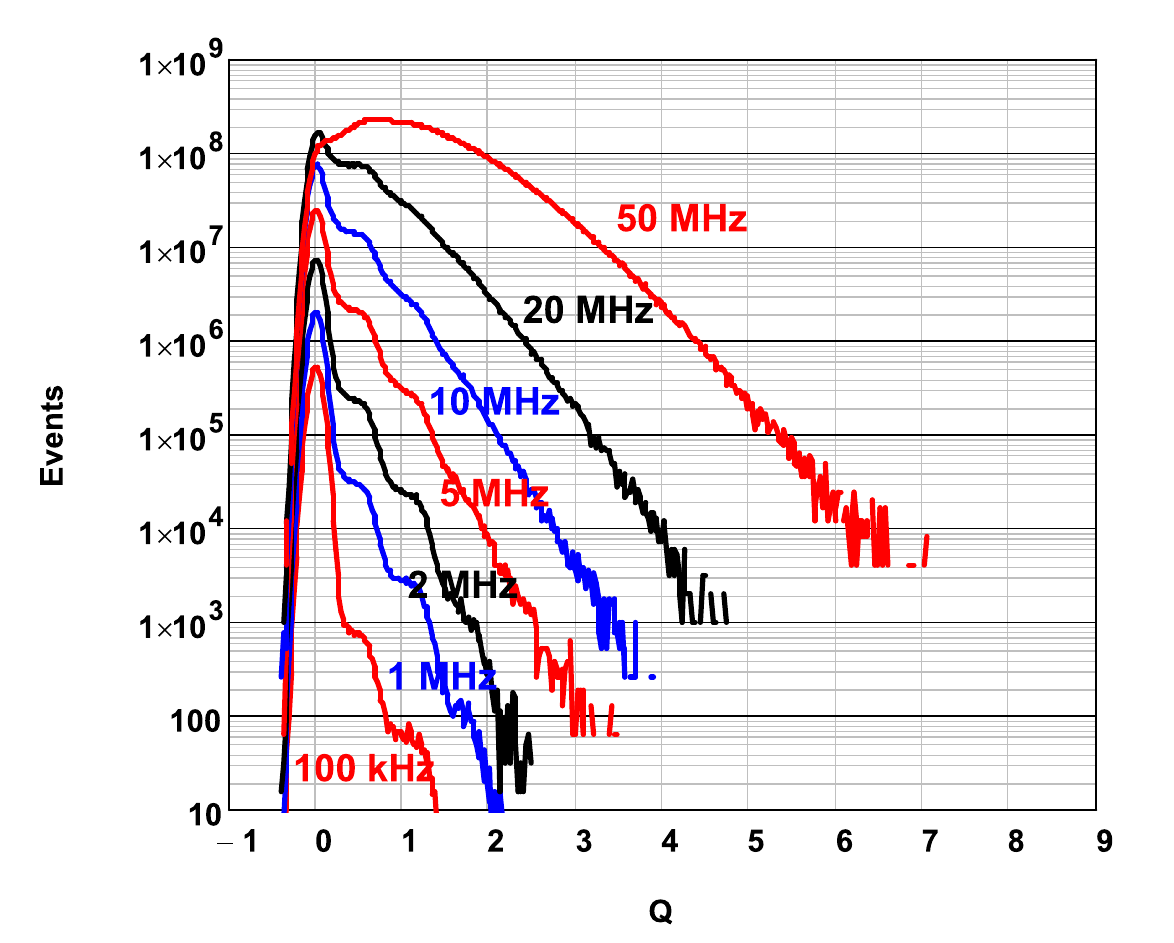}
    \caption{ }
    \label{fig:DCDCR20ns}
    \end{subfigure}%
   \caption{Same as Fig.~\ref{fig:QvsDCR100ns} with $t_{gate} = 20$~ns.}
   \label{fig:QvsDCR20ns}
  \end{figure}

 Figs.~\ref{fig:QvsDCR100ns} and \ref{fig:QvsDCR20ns} show for gate lengths $t_{gate} = 100$~ns and 20~ns simulated charge spectra with and without illumination for $DCR$~values between 100~kHz and 50~MHz.
 For the spectra with illumination the number of light-induced primary Geiger discharges $N_{\gamma G} = 5$.
 For the pulse shape a single exponential with the time constant $\tau _s = 20$~ns, and for the probabilities of after-pulses, and of prompt and delayed cross-talk the values 15~\%, 10~\% and 10~\% are assumed.

Up to $DCR = 1$~MHz, which corresponds to a  probability of 1~\% of a dark count in the 100~ns gate, the charge spectra with illumination are hardly affected.
 With increasing $DCR$ the mean  charge and the background below the peaks increase, and the amplitudes of the individual peaks decrease.
 For $DCR = 50$~MHz, which corresponds to an average of 5 dark counts in a 100~ns gate, the individual peaks have essentially disappeared and the mean charge is shifted by approximately
 $DCR \cdot t_{gate} \cdot  \int _0 ^{t_{gate}} I_1 (t)~\mathrm{d}t $.
 Comparing the charge spectra with illumination for $t_{gate} = 100$~ns (Fig.~\ref{fig:QDCR100ns}) with the ones for $t_{gate} = 20$~ns  (Fig.~\ref{fig:QDCR20ns}) shows that the degradation of the spectra with $DCR$ is worse for the longer $t_{gate}$.
 For $t_{gate} = 20$~ns peaks corresponding to different number of Geiger discharges are still visible at 50~MHz.
 This reason is that more dark counts are integrated for the longer gate.

 Similar effects are observed when comparing the dark spectra for $t_{gate} = 100$~ns (Fig.~\ref{fig:DCDCR100ns}) and 20~ns (Fig.~\ref{fig:DCDCR20ns}):
 The longer the gate the more dark counts are integrated which results in an increase of the mean charge.
 It can also be noted that for $t_{gate} = 100$~ns up to 20~MHz pedestal and single Geiger discharge peak are well separated, and $DCR$ can be determined using Eq.~\ref{eq:DCR}.




 \section{Summary and outlook}
  \label{sect:Summary}

 A relatively simple and flexible Monte Carlo program for the simulation of the response of SiPMs is presented.
 For a given dark-count rate and the mean number, frequency and time distribution of primary Geiger discharges from a light source, correlated Geiger discharges corresponding to after-pulses and prompt and delayed cross-talk are generated.
 A number of different physics-based models and statistical treatments for the simulation of correlated Geiger discharges are implemented and can be selected.

 To illustrate the usefulness of the simulation program, a number of examples for its use are given:
 \begin{itemize}
   \item Simulation of amplitude versus time difference distributions of consecutive Geiger discharges for dark counts.
   \item Comparison of the current transients measured with a transient recorder with simulated transients.
   \item Comparison of the charge spectra measured with a gated QDC (charge to digital convertor) with simulated spectra.
   \item Simulation of the influence of a fast component of the SiPM pulses on the charge spectra as recorded with a QDC.
   \item Simulation of the influence of dark-count rates up to values of 50~MHz on the charge spectra as recorded with a QDC with different gate lengths.
 \end{itemize}
 These examples demonstrate that such a simple but flexible simulation program can be used to
 test and verify analysis programs used to extract SiPM parameters from experimental data,
 determine SiPM parameters by comparing measurement and simulation results, and
 evaluate and understand the impact on the measurement of photons of different SiPM parameters, probability distributions for cross-talk and after-pulses and readout schemes.

  \section*{Bibliography}
   \label{sect:Bibliography}

\end{document}